\documentclass[a4paper, superscriptaddress, amsfonts, amssymb, amsmath, reprint, showkeys, nofootinbib, twoside]{revtex4-1}

\usepackage{graphicx}
\usepackage{dcolumn}
\usepackage{bm}
\usepackage{siunitx}
\usepackage{braket}

\usepackage{color}
\usepackage{xcolor}
\usepackage{multirow}
\usepackage{amsmath}

\usepackage{hhline}
\usepackage{threeparttable}

\def\U#1{{\rm #1}} 
\def\u#1{_{\rm #1}}

\newcommand{\expect}[1]{\langle #1 \rangle} 
\newcommand{\od}[2]{\frac{\mathrm{d} #1}{\mathrm{d} #2}}

\begin{document}


\title{Low-noise quantum frequency conversion with cavity enhancement of converted mode}

\author{Shoichi Murakami}
 \email{smurakami@qi.mp.es.osaka-u.ac.jp}
 \affiliation{ Graduate School of Engineering Science, Osaka University, Osaka 560-8531, Japan}%
  \affiliation{ Center for Quantum Information and Quantum Biology, Osaka University, Osaka 560-0043, Japan}%
 
\author{Toshiki Kobayashi}%
  \affiliation{ Graduate School of Engineering Science, Osaka University, Osaka 560-8531, Japan}%
  \affiliation{ Center for Quantum Information and Quantum Biology, Osaka University, Osaka 560-0043, Japan}%

\author{Shigehito Miki}
 \affiliation{ Advanced ICT Research Institute, National Institute of Information and Communications Technology (NICT), Kobe, 651-2492, Japan}%
\author{Hirotaka Terai}
 \affiliation{ Advanced ICT Research Institute, National Institute of Information and Communications Technology (NICT), Kobe, 651-2492, Japan}%

\author{Tsuyoshi Kodama}
 \affiliation{ Advanced ICT Research Institute, National Institute of Information and Communications Technology (NICT), Kobe, 651-2492, Japan}%
 \affiliation{ Hamamatsu Photonics K. K., Shizuoka, Japan}

\author{Tsuneaki Sawaya}
 \affiliation{ Hamamatsu Photonics K. K., Shizuoka, Japan}
 
\author{Akihiko Ohtomo}
 \affiliation{ Hamamatsu Photonics K. K., Shizuoka, Japan}

\author{Hideki Shimoi}
 \affiliation{ Hamamatsu Photonics K. K., Shizuoka, Japan}

\author{\\Takashi Yamamoto}%
  \affiliation{ Graduate School of Engineering Science, Osaka University, Osaka 560-8531, Japan}%
  \affiliation{ Center for Quantum Information and Quantum Biology, Osaka University, Osaka 560-0043, Japan}%

\author{Rikizo Ikuta}%
  \affiliation{ Graduate School of Engineering Science, Osaka University, Osaka 560-8531, Japan}%
  \affiliation{ Center for Quantum Information and Quantum Biology, Osaka University, Osaka 560-0043, Japan}%

\date{\today}

\begin{abstract}
Quantum frequency conversion (QFC) 
which converts the frequencies of photons while preserving the quantum state 
is an essential technology for realizing the quantum internet and quantum interconnect. 
For the QFC based on the frequency downconversion 
from visible to the telecom wavelengths around \SI{1.5}{\micro m}, 
it is widely known that noise photons produced by the strong pump light used for QFC 
contaminate the frequency-converted photon, 
which degrades the quality of the quantum property of the photon after QFC. 
In conventional QFC experiments, noise photons are removed using external narrowband frequency filter systems. 
In contrast, in this study, 
we implement a compact QFC device integrating the cavity structure only for the converted mode. 
While the cavity structure can enhance not only the desired QFC efficiency but also the noise photon generation rate, 
we show that the cavity-enhanced QFC followed 
by a relatively wide bandpass filter achieves the signal-to-noise ratio
comparable to the QFCs with external narrowband filters. 
We experimentally demonstrate the cavity-enhanced QFC using a single photon at \SI{780}{nm} to \SI{1540}{nm}, in which the non-classical photon statistics is clearly observed after QFC. 
\end{abstract}

\maketitle
\section{Introduction}
In recent years, the development of quantum computers has been remarkably advancing across various quantum matter systems~\cite{de2021materials,bluvstein2024logical}.
Consequently, quantum networks connecting these quantum matter systems are actively explored~\cite{kimble2008quantum,wehner2018quantum,wei2022towards}.
While each medium has its own affinity for specific wavelengths of photons, long-distance communication via optical fibers only allows for photons at the telecom bands with low attenuations of \SI{0.2}{dB/km} around \SI{1550}{nm} and \SI{0.3}{dB/km} around \SI{1310}{nm}. 
The frequency mismatch necessitates a quantum frequency conversion~(QFC)~\cite{kumar1990quantum} which converts the frequencies of the photons to the telecom wavelengths while preserving its quantum state~\cite{ikuta2011wide}. 
So far, numerous QFC experiments converting to the telecom wavelengths
have been conducted on photons emitted from various quantum matter systems, such as neutral atoms~\cite{ikuta2018polarization,van2020long,yu2020entanglement,van2022entangling,luo2022postselected,liu2024creation,zhou2024long}, ions~\cite{bock2018high,krutyanskiy2019light}, NV centers~\cite{tchebotareva2019entanglement}, SiV centers~\cite{bersin2024telecom,knaut2024entanglement}, and so on.

\begin{figure}[t]
\centering\includegraphics[width=\linewidth]{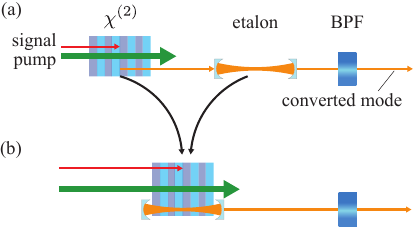}
\caption{(a) Conventional QFC system and (b) our QFC system}
\label{fig:concept}
\end{figure}

An important challenge common to all the QFCs is reducing background noise photons derived from the Raman scattering and/or spontaneous parametric downconversion~(SPDC) 
produced by the high-power pump light used for QFC~\cite{pelc2011long,ikuta2014frequency,maring2018quantum,Strassmann:19}.
In conventional QFCs using continuous-wave~(cw) pump lights, the noise photons are generated over a broad temporal and spectral range. Therefore, it is important to properly filter in both time and frequency domains, especially in QFCs that convert narrow linewidth photons emitted from quantum matters, where narrowband frequency filtering becomes critical.
For this, a frequency filtering system composed of a narrowband etalon followed by a relatively wider bandpass filter~(BPF) shown in Fig.~\ref{fig:concept}~(a) has been developed in recent QFC experiments~\cite{ikuta2018polarization,van2020long,van2022entangling,zhou2024long,krutyanskiy2019light,walker2018long,stolk2022telecom}. 
The narrowband etalon filter requires additional optical components for mode matching and the frequency stabilization mechanism, which will complicate the overall QFC system and introduce additional losses.

In this paper, 
we demonstrate QFC based on a periodically poled lithium niobate waveguide resonator~(PPLN-WR), 
which integrates the etalon filter into the QFC as shown in Fig.~\ref{fig:concept}~(b). 
The PPLN-WR confines only the converted photon generated by QFC into the cavity 
but does not confine the signal photon and pump light. 
This configuration avoids the severe mode matching of light to the cavity and does not induce temperature instability in the PPLN-WR~\cite{ikuta2022optical}. 
While the cavity structure of the PPLN-WR can enhance not only the desired QFC process 
but also unwanted noise photon generation, 
we theoretically show that by a resonator with a practically feasible finesse, 
a higher SNR is achieved compared with conventional QFCs 
considering the use of the BPF after QFC with its bandwidth corresponding to the FSR of the resonator. 
We experimentally demonstrate the cavity-enhanced QFC of the single photon from \SI{780}{nm} to \SI{1540}{nm} using a pump light at \SI{1581}{nm}. 
The non-classical photon statistics is clearly shown after QFC. 
Together with the observed spectral properties and pump power dependency of the noise photons, we see the cavity-enhanced QFC works well as theoretically predicted. 

\section{Basic concept for the SNR improvement using cavity-enhanced QFC}
\label{sec:theory}
\begin{figure}[t]
\centering\includegraphics[width=\linewidth]{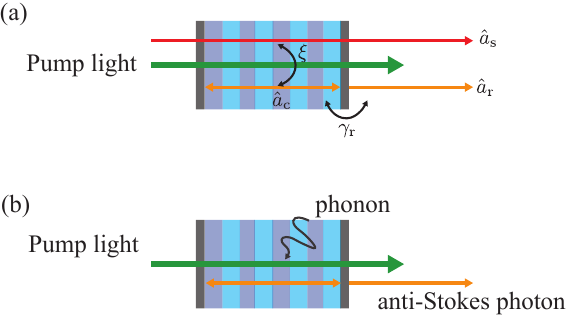}
\caption{Theoretical model of (a) the frequency conversion in the cavity and (b) the generation of AS photons.}
\label{fig:theoretical model}
\end{figure}

\begin{figure}[t]
\centering\includegraphics[width=7.2cm]{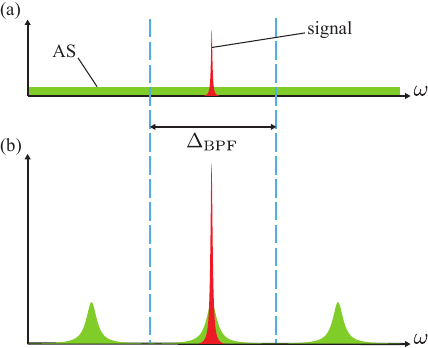}
\caption{The spectrum of signal and AS photons (a) without and (b) with a cavity. 
The total amounts of AS photons within the range of FSR are the same 
when the pump powers used for both cases are equal. }
\label{fig:AS}
\end{figure}
We first review the theory of difference frequency generation~(DFG) 
with a cavity structure for the converted mode~\cite{ikuta2022optical}. 
We assume the signal photon is in a single frequency mode at angular frequency $\omega\u{s}$, 
and it is frequency-converted through DFG
using a sufficiently strong pump light at angular frequency $\omega\u{p}$. 
The converted frequency is $\omega\u{r}=\omega\u{s}-\omega\u{p}$. 
In the model, 
around the cavity resonant mode at $\omega\u{cav}$
is coupled to two external modes as shown in Fig.~\ref{fig:theoretical model}~(a). 
One of the external modes is the signal mode at $\omega\u{s}$ of DFG. 
The coupling constant $\xi$ is proportional to the complex amplitude of the pump light.
The other external mode is the converted mode at $\omega\u{r}$
outside the cavity. 
The coupling constant $\sqrt{\gamma_\mathrm{r}}$ is 
related to the reflectance of the end mirror of the cavity. 
The complex amplitudes of unconverted and conversion efficiencies are written as 
\begin{align}\label{trans_prob}
t_\mathrm{ss} &= \frac{\frac{1}{2}(1-\tilde{C})-i\tilde{\Delta}_\mathrm{c}}{\frac{1}{2}(1+\tilde{C})-i\tilde{\Delta}_\mathrm{c}},\\
\label{ref_prob}
r_\mathrm{rs} &= \sqrt{\tilde{\gamma}_\mathrm{r}}\frac{e^{-i\phi}\sqrt{\tilde{C}}}{\frac{1}{2}(1+\tilde{C})-i\tilde{\Delta}_\mathrm{c}},
\end{align}
where $\phi$ is the phase of pump light, $\tilde{\gamma}_\mathrm{r} = \gamma_\mathrm{r}/\gamma_\mathrm{all}$, 
$\tilde{C} = |\xi|^2 / \gamma_\mathrm{all}$ 
and $\tilde{\Delta}_\mathrm{c}=(\omega\u{r}-\omega\u{cav})/\gamma_\mathrm{all}$.  
$\gamma_\mathrm{all} = \gamma_\mathrm{r} + \gamma_\mathrm{int}$ is the total loss 
determined by $\gamma_\mathrm{r}$ and internal loss of the cavity $\gamma_\mathrm{int}$. 
$\tilde{C}$ is proportional to the pump power $P$, 
and can be described by $\tilde{C} = \tilde{\alpha}P$ 
using a proportional coefficient $\tilde{\alpha}$. 
The maximum conversion efficiency is achieved at $P=\tilde{\alpha}^{-1}$. 
The value of $\tilde{\alpha}$ is related to the cavity enhancement factor, 
which we will discuss later. 

In the QFC experiments, it is known that the strong pump light is used not only for DFG of the signal photon but also for other unwanted nonlinear optical processes in which single-photon-level noise photons are produced.
In the QFCs where the converted frequency and the pump frequency are close to each other 
with the condition of $\omega\u{s}/2 > \omega\u{p}$, 
the anti-Stokes~(AS) photons produced by the pump light 
contaminate the converted photon as the noise source~\cite{pelc2011long}. 
The AS photons have higher frequencies than the pump light because they receive the energies from phonons in the material. 
We treat the process as the sum frequency generation~(SFG) of 
the phonons to frequencies around the converted mode by the pump light. 
Due to the cavity structure for the converted mode, the AS photons are inside the cavity modes, whereas the phonons are not confined to the cavity. 
We thus regard the AS photon generation process as the SFG process in the cavity resonant to the converted mode, which is described in Fig.~\ref{fig:theoretical model}~(b).
Fig.~\ref{fig:AS} shows the schematic drawing of the spectra of AS photons with and without cavity structure. 
While the AS photons are gathered into the resonant peaks with the cavity enhancement, the generated AS photons inside the cavity become 
half the amount of AS photons without cavity at the pump power equal to each other when the bandwidth $\Delta_\mathrm{BPF}$ of the photon detection is equal to the FSR of the cavity. The detailed calculation is described in Appendix~\ref{app:ASphoton}. 

Based on the above treatment of the noise photons,  
we compare the SNR of our cavity-enhanced QFC with that of the conventional QFC without cavities 
for the detection bandwidth satisfying $\delta\u{FWHM}\ll \Delta_\mathrm{BPF}\leq \delta\u{FSR}$, 
where $\delta\u{FWHM}$ and $\delta\u{FSR}$ are the full width at the half maximum~(FWHM) and FSR of the cavity. 
We assume that 
the linewidth of the signal photon is sufficiently narrower than $\delta\u{FWHM}$, 
such that the conversion efficiency is described by 
$\eta\u{cav}=|r\u{rs}|^2
=4\tilde{\gamma}_\mathrm{r}\tilde{\alpha}P(1+\tilde{\alpha}P)^{-2}$
using Eq.~(\ref{ref_prob}) with $\tilde{\Delta}_\mathrm{c}=0$. 
For the noise photons, 
using the extraction efficiency from the cavity of $\tilde{\gamma}\u{r}$, 
we describe the amount of noise photons by 
$N\u{noise,cav} = \tilde{\gamma}\u{r}\alpha\u{noise} P/2$, 
where $\alpha\u{noise}$ is the proportionality constant of the AS photons without the cavity~(see Appendix~\ref{app:ASphoton}).
As a result, the SNR of the cavity-enhanced QFC is described by 
\begin{align}
SNR\u{cav} = \frac{\eta\u{cav}}{N\u{noise,cav}}
= \frac{8\tilde{\alpha}}{\alpha\u{noise}(1+\tilde{\alpha}P)^2}. 
\label{SNRc}
\end{align} 
For the conventional QFC without a cavity structure, the conversion efficiency 
is described by $\eta\u{nocav}=\sin^2(\sqrt{BP})$ using a constant $B$. 
The amount of the AS photons is described by 
$N\u{noise,nocav} = \alpha\u{noise} P \Delta_\mathrm{BPF}/\delta\u{FSR}$. 
Thus, the SNR is given by 
\begin{align}
SNR\u{nocav}=\frac{\eta\u{nocav}}{N\u{noise,nocav}}
= \frac{B\U{sinc}^2(\sqrt{BP})}{\alpha\u{noise} \Delta_\mathrm{BPF}/\delta\u{FSR}}. 
\label{SNR}
\end{align}
\begin{figure}[t]
\centering \scalebox{1}{\includegraphics{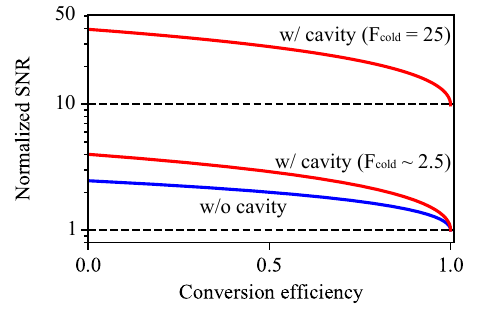}}
\caption{Normalized SNR v.s. conversion efficiency. 
The normalization is based on taking the value 
of $SNR\u{nocav}$ at $\eta\u{nocav}=1$ 
by using $B/\alpha\u{noise}=\pi^2/4\sim 2.5$. 
Blue: conventional QFC. 
Red: cavity-enhanced QFC with $F\u{cold}=8/\pi\sim 2.5$ and $25$. 
}
\label{fig:SNR}
\end{figure}
The SNR of the conventional QFC is improved by using a narrower BPF, 
whereas $SNR\u{cav}$ is independent of $\Delta\u{BPF}(\gg \delta\u{FWHM})$
because most of the AS photons are inside within the range of $\delta\u{FWHM}$. 
We consider the case of $\Delta\u{BPF}=\delta\u{FSR}$ in the following discussion. 

The conversion efficiency of the QFC with the cavity 
is enhanced by the factor of $F_\mathrm{cold}/\pi$ 
for the converted mode inside the cavity in the low pump power regime~\cite{berger1997second,liscidini2006second}, 
where $F_\mathrm{cold}$ is the finesse of the cavity without QFC. 
From 
$\eta\u{cav}=4\tilde{\gamma}\u{r}\tilde{\alpha}P+\mathcal{O}(P^2)$ and 
$\eta\u{nocav}=BP+\mathcal{O}(P^2)$ for $P\ll 1$, 
we obtain $\tilde{\alpha}=F\u{cold}B/4\pi$. 
Together with $\eta\u{cav}$, $\eta\u{nocav}$, Eqs.~(\ref{SNRc}) and (\ref{SNR}), 
the dependencies of the SNRs on conversion efficiencies are 
calculated for the QFCs with and without cavity structure. 
The results with the use of $\Delta\u{BPF}=\delta\u{FSR}$ are shown in Fig.~\ref{fig:SNR}, 
in which the SNRs are normalized by $SNR\u{nocav}$ at $\sqrt{BP}=\pi/2$ 
by using $B/\alpha\u{noise}=\pi^2/4$. 
To obtain higher SNRs for any conversion efficiency using the cavity enhancement, $F\u{cold}\geq 8/\pi\sim 2.5$ is required. 
For $F\u{cold} = 25$, the SNR improves by one order of magnitude. 

\section{Experiments}
\subsection{Experimental setup}
\begin{figure*}[t]
\centering\includegraphics{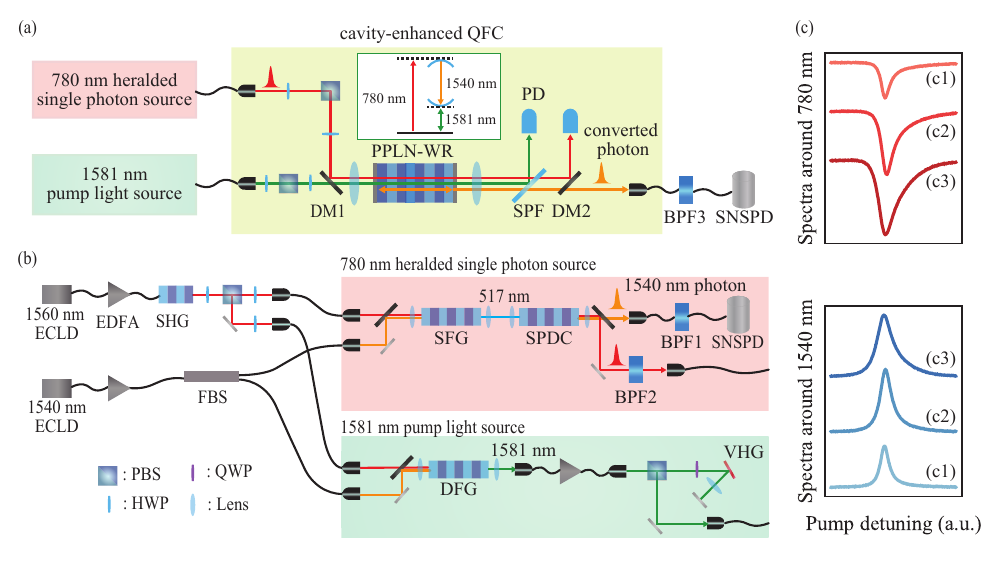}
\caption{Experimental setup (a) for the QFC experiment and (b) for the preparation of 780 nm heralded single photon source and 1581 nm pump light source.
HWP: half-wave plate, PBS: polarizing beamsplitter, DMs: dichroic mirrors, SPF: short pass filter, BPFs: bandpass filters, SNSPDs: superconducting nanostrip single-photon detectors, 
ECLDs: external cavity laser diodes, EDFAs: erbium-doped fiber amplifiers,  FBS: fiber beamsplitter, QWP: quarter-wave plate, and VHG: volume holographic grating. (c) Observed two spectra of transmitted light and converted light for the pump power of (c1) 33.3 mW, (c2) 94.0 mW, and (c3) 148 mW.}
\label{setup}
\end{figure*}
The experimental setup for the cavity-enhanced QFC 
from \SI{780}{nm} to \SI{1540}{nm} is shown in Fig.~\ref{setup}~(a). 
The details of the preparation setups 
for the \SI{780}{nm} heralded single photon and 
\SI{1581}{nm} pump light for the QFC are shown in Fig.~\ref{setup}~(b). 
We used two continuous wave~(cw) lights at \SI{1560}{nm} and \SI{1540}{nm} 
emitted from two external cavity laser diodes~(ECLDs) 
after amplification by erbium-doped fiber amplifiers~(EDFAs). 
The \SI{1560}{nm} laser was frequency doubled by second harmonic generation~(SHG).
The SHG light at \SI{780}{nm} and the \SI{1540}{nm} light produced
the \SI{517}{nm} light by sum frequency generation~(SFG), 
which was used as a pump light to generate a photon pair 
at \SI{780}{nm} and \SI{1540}{nm} through the SPDC process. 
The \SI{1540}{nm} photon was detected using a superconducting nanostrip single photon detector~(SNSPD) 
after BPF1 with the bandwidth of \SI{0.03}{nm}, and then the \SI{780}{nm} photon used for QFC was heralded. 
The heralded photon passed through BPF2 with the bandwidth of \SI{0.4}{nm} 
and then was coupled to a polarization-maintaining fiber~(PMF). 
On the other hand, the \SI{1581}{nm} pump light for QFC was prepared 
by DFG using the SHG light at \SI{780}{nm} and \SI{1540}{nm} light. 
After the amplification at the maximum power of around \SI{250}{mW}, 
the \SI{1581}{nm} light passed through a volume holographic grating~(VHG) with the bandwidth of \SI{1}{nm} twice 
to remove noise photons such as amplified spontaneous emission~(ASE) photons. 
For all of the above nonlinear optical processes of SHG, SFG, DFG, and SPDC, 
we used conventional PPLN waveguides without cavity structure. 

For the QFC in Fig.~\ref{setup}~(a), 
the heralded single photon at \SI{780}{nm} and the pump light at \SI{1581}{nm} 
were combined at a dichroic mirror~(DM1), and then they were coupled to the PPLN-WR. 
The PPLN-WR in the experiment satisfies the type-0 quasi-phase-matching condition. 
Both end faces of the waveguide are flat polished, and coated by dielectric multilayers. 
The length of the waveguide is $L^\U{ex}=\SI{13.26}{mm}$ 
which corresponds to the FSR of $\delta\u{FSR}=\SI{5.2}{GHz}$ for \SI{1540}{nm}~(see Appendix~\ref{app:A}). 
The reflectances of the front and rear faces for \SI{1540}{nm} are 
asymmetrically designed to extract the converted photon from the rear side efficiently. 
We show the reflectances of the end faces in Appendix~\ref{app:B} 
including those for the signal photon and the pump light. 

After the QFC, the pump light and signal photon were separated from the converted photon by a short-pass filter~(SPF) and DM2, respectively. 
The \SI{1540}{nm} photon was coupled to a PMF, 
and frequency filtered by BPF3 with the bandwidth of \SI{0.03}{nm} 
corresponding to \SI{73}{\%} of the FSR. 
Finally, it was detected by SNSPD. 
All SNSPDs used in the experiment were developed by Hamamatsu Photonics and NICT.
The quantum efficiencies of the SNSPDs used in the experiment are $\sim$ 0.8. 
Including the detection efficiency, 
the transmittance of the optical circuit for the converted photon was about 
$T\u{circ}=0.08$. 
We collected the coincidence events using a time-to-digital converter 
with electrical signals coming from SNSPDs. 

\subsection{Evaluation of the cavity enhancement effect}
\begin{figure}[h]
\centering\includegraphics{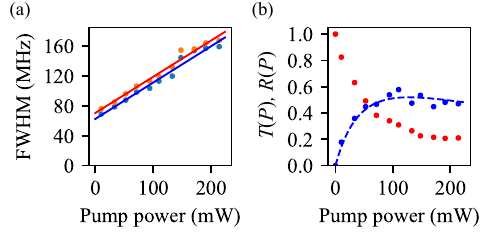}
\caption{(a) Pump power dependencies of observed bandwidths of transmitted (red) and converted (blue) light. (b) Pump power dependencies of $T(P) = |t_\mathrm{ss}|^2$~(red) and $R(P) = |r_\mathrm{rs}|^2$~(blue) for  $\tilde{\Delta} = 0$. The blue dashed curve was obtained from Eq.~(\ref{ref_prob}) using $(\tilde{\alpha}^\U{ex})^{-1}\sim$ \SI{144}{mW} estimated from Fig.~\ref{fig:fwhm_and_conv}~(a). 
The maximum conversion efficiency is determined by a factor of $\tilde{\gamma}\u{r} = 0.7$ calculated from the result of Ref.~\cite{ikuta2022optical}.}
\label{fig:fwhm_and_conv}
\end{figure}
As a preliminary experiment, we evaluated the effect of the cavity enhancement on the DFG from \SI{780}{nm} to \SI{1540}{nm} in the PPLN-WR using the laser light at \SI{780}{nm}. 
We measured the transmission spectra at \SI{780}{nm} and \SI{1540}{nm} 
after DFG for several pump powers. 
An example of the spectra for a pump power is shown in  Figs.~\ref{setup}~(c). 
As expected from Eqs.~(\ref{trans_prob}) and (\ref{ref_prob}), 
we observed the broadening of the spectral bandwidths as the pump power increased. 
The pump power dependencies of the FWHMs 
are shown in Figs.~\ref{fig:fwhm_and_conv}~(b). 
By fitting the experimental results with a function 
$\alpha^\U{ex} P + \gamma_\mathrm{all}^\mathrm{ex}$, 
we obtained $\alpha^\U{ex} = \SI{0.49}{MHz/mW}$ and 
$\gamma_\mathrm{all}^\mathrm{ex} = \SI{70.4}{MHz}$. 
The finesse without the pump light is estimated to be 
$F\u{cold}^\U{ex}=\delta\u{FSR}/\gamma_\mathrm{all}^\mathrm{ex} = 74$. 

From the result of Fig.~\ref{fig:fwhm_and_conv}~(a), 
we obtained the pump power dependencies of the internal conversion efficiencies 
as shown in Fig.~\ref{fig:fwhm_and_conv}~(b). 
The behavior is in good agreement with the theoretically predicted curves in Eq.~(\ref{ref_prob}) with the use of parameters 
$\alpha^\U{ex}$ and $\gamma_\mathrm{all}^\mathrm{ex}$
estimated by Figs.~\ref{fig:fwhm_and_conv}~(c). 
The maximum conversion efficiency was achieved 
at $(\tilde{\alpha}^{\U{ex}})^{-1}=\gamma\u{all}^\U{ex}/\alpha^\U{ex} \sim $ \SI{144}{mW}.
By borrowing the reported value of $B^\U{ref}=\SI{17.3e-3}{\mbox{/}mW}$
using a PPLN waveguide~(the length $L^\U{ref}=\SI{45}{mm}$) without cavity structure~\cite{murakami2023quantum} and considering 
the difference in the lengths of the waveguides, 
the enhancement factor is estimated to be 
$4\tilde{\alpha}^\U{ex}/B^\U{ref} (L^\U{ref}/L)^{2} = 18$. 
The value is not significantly different from $F\u{cold}^\U{ex}/\pi \sim 23$. 
From the results, the singly resonant structure for the converted mode works well 
for the cavity-enhanced frequency conversion process. 
We observed the larger enhancement effect for the frequency conversion 
to \SI{1522}{nm} using the \SI{1600}{nm} pump light. 
The result is shown in Appendix~\ref{app:C}. 

\subsection{Properties of AS photons and QFC experiment using the single photon}
\begin{figure*}[t]
\centering\includegraphics{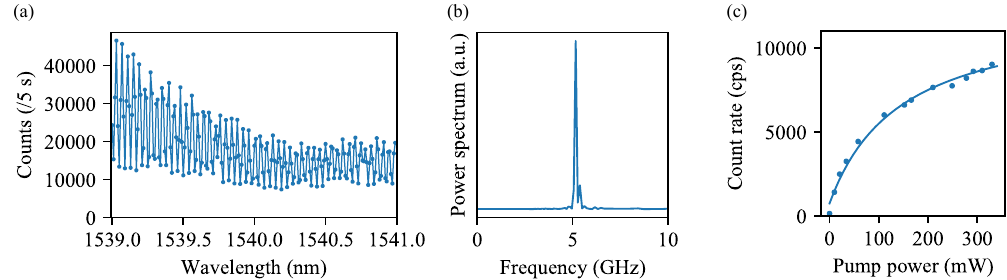}
\caption{(a) Photon count distribution of AS photon around 1540 nm. 
(b) The power spectrum of AS photons obtained by performing the Fourier transform to the data (a). (c) The pump power dependency of the number of AS photons. The blue line is calculated using Eq.~(\ref{eq:noise_pump}) and the data of bandwidths of \SI{1540}{nm} converted light.} 
\label{fig:spectrum}
\end{figure*}
We investigated the properties of the AS photons 
and then perform the QFC experiment. 
We first measured the single photon counts of the AS photons 
by scanning the center wavelengths of BPF1 over \SI{2}{nm} 
in \SI{0.01}{nm} increments with the minimal bandwidth of \SI{0.03}{nm}. 
The result is shown in Fig.~\ref{fig:spectrum}~(a), 
in which a clear oscillation of the photon counts is observed. 
From the Fourier transform to the data shown in Fig.~\ref{fig:spectrum}~(b), 
the period of the peaks was found to be $\SI{5.2}{GHz}\pm \SI{0.1}{GHz}$.
This result matches the FSR of our PPLN-WR well. 
This implies the spectrum of generated AS photons resonates to the PPLN-WR.

Next, we measured the pump power dependency of the AS photons. 
The bandwidth of the BPF was set to be \SI{0.03}{nm}. 
The result is shown in Fig.~\ref{fig:spectrum}~(c). 
While there is the linear dependency in the lower pump power region, 
a slower increase of the noise photons is observed for a larger pump power,
which indicates the frequency up-conversion process of the generated AS photons as in Ref.~\cite{maring2018quantum}. 
We fitted the experimental data using Eq.~(\ref{eq:noise_pump}) in Appendix \ref{app:ASphoton}, which is obtained by modeling the joint process of the generation and the frequency up-conversion of the AS photons inside the cavity.
From the result and 
$\tilde{\gamma}\u{r} = 0.7$ derived from Ref.~\cite{ikuta2022optical} with the assumption that $\gamma\u{int}$ is proportional to the cavity length, 
we obtained $\alpha_\mathrm{noise}^\U{ex} = \SI{230}{cps/\milli\watt}$.
Because of $\Delta_\mathrm{BPF} = \SI{3.79}{\giga\Hz}\gg \delta\u{FWHM}$ 
in this experiment, the proportional coefficient $\alpha_\mathrm{noise}^\U{ex}$ 
should not show the cavity enhancement effect as we described in Sec.~\ref{sec:theory}. 
To see this, we compare 
the reported value $\alpha\u{noise}^\U{ref} =\SI{970}{cps/\milli\watt}$ 
in Ref.~\cite{murakami2023quantum}.
By normalizing the length of the PPLN and the bandwidth, 
the proportional coefficients just after the PPLNs are estimated to be 
$\alpha\u{noise}/(L \Delta\u{BPF} T\u{circ})\sim$
\SI{60}{cps\per(\milli\meter\,\giga\Hz\,\milli W)}
and 
$\alpha\u{noise}^\U{ref}/(L^\U{ref}\Delta\u{BPF}^{\U{ref}}T\u{circ}^\U{ref})\sim $
\SI{25}{cps\per(\milli\meter\,\giga\Hz\,\milli W)}, 
where $\Delta\u{BPF}^\U{ref} = \SI{12.6}{\giga\Hz}$ 
and $T\u{circ}^\U{ref}=0.09$. 
We see the values are surely consistent with the theory in Sec.~\ref{sec:theory}. 
Similar experiments for the AS photons around \SI{1522}{nm}
were performed as shown in Appendix~\ref{app:C}. 

\begin{figure}[t]
\centering\includegraphics[width=\linewidth]{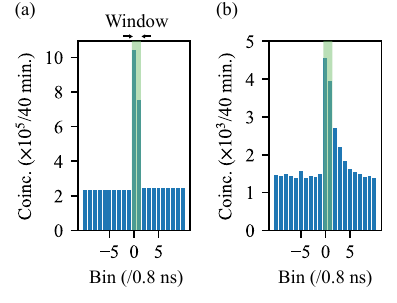}
\caption{(a) Observed coincidence counts in \SI{40}{minutes} between the signal and idler photons. (b) Observed coincidence counts in \SI{40}{minutes} between the converted and idler photons with $P\sim$ \SI{140}{mW}. 
The time resolution and width of the window are \SI{0.8}{ns} and \SI{1.6}{ns}.}
\label{fig:histogram}
\end{figure}
Finally, we performed the QFC experiment using the single photon at \SI{780}{nm}
heralded by the \SI{1540}{nm} photon produced by the SPDC. 
The coincidence counts between the SPDC photon pairs 
without QFC are shown in Fig.~\ref{fig:histogram}~(a). 
The cross-correlation function within the coincidence time window of \SI{1.6}{ns} is $g^{(2)}\u{in}=3.819\pm 0.003$. 
After the QFC with the pump power of $\sim$ \SI{140}{mW}, 
the experimental result of 
the coincidence counts between the heralding photon and the converted photon at \SI{1540}{nm} is shown in Fig.~\ref{fig:histogram}~(b).
The cross-correlation function of the photon pair is estimated to be 
$g^{(2)}\u{out}=2.94\pm 0.0 3$, 
which surpasses the classical limit of 2. 
We successfully demonstrated the QFC 
with the cavity structure for the converted mode. 

\section{Discussion}
In the QFC experiment, we used the heralded single-photon 
produced by the SPDC process. 
The linewidth of the signal photon at \SI{780}{nm} is about \SI{3.8}{GHz} 
corresponding to $0.73 \delta\u{FSR}$. 
The broad linewidth results in significantly lower efficiency, 
which is estimated to be 0.07. 
Nonetheless, 
the successful QFC was achieved without a significant reduction 
in the cross-correlation function. 
From the equation of 
$g^{(2)}\u{out}=(g^{(2)}\u{in}\zeta + 1)/(\zeta +1)$ in Ref.~\cite{albrecht2014waveguide,ikuta2016} 
where $\zeta$ is the intensity ratio of the signal photon to 
the equivalent input noise to the converter, 
$\zeta =2.1$ is estimated. 
From the enhancement factor of 18 obtained in the previous section, 
the intensity correlation function after QFC will be 1.3 
at the maximum conversion efficiency if the converter has no cavity structure. 
As a result, we conclude that the non-classicality of 
the cross-correlation function after QFC is due to the cavity enhancement effect on the converted mode. 

\begin{table}
\centering
\begin{tabular}{|c||ccc|}\hline
   & \multicolumn{3}{c|}{cavity confinement}\\ \cline{2-4}
   &  no cavity & converted mode & signal mode \\ \hline\hline
    $\Delta\u{BPF}=\delta\u{FSR}$ & 1 & $2F_\mathrm{c} / \pi$ &  $F_\mathrm{s} / \pi$  \\
    $\Delta\u{BPF}=\delta\u{FWHM}$ & $F_\mathrm{c}$ & $2F_\mathrm{c} / \pi$ & $F_\mathrm{c}F_\mathrm{s} / \pi$\\\hline
        \end{tabular}
    \caption{Normalized SNRs of QFCs in various configurations. 
    }
    \label{tbl:SNR}
\end{table}
We discuss the dependencies of the SNR improvement of QFC 
on the cavity enhancement and detection bandwidth. 
In the section, we denote the finesse of $F\u{cold}$ by $F\u{c}$ 
to clearly express that the factor comes from the confinement of the converted~(c) mode. 
In our demonstration, we showed the SNR improvement of QFC 
with the confinement of the converted mode 
when the detection bandwidth is $2\Delta\u{BPF}=\delta\u{FSR}$. 
The theoretical limit of the improvement factor is $F\u{c}/\pi$ 
compared with the case of no cavity structure. 
If the narrower BPF of $\Delta\u{BPF}=\delta\u{FWHM}$ is used 
in the conventional QFC without cavities, 
the improvement factor of $F\u{c}$ is expected. 
We consider the SNR of QFC 
with the cavity for the converted mode followed by the BPF 
with the bandwidth of $\Delta\u{BPF}=\delta\u{FWHM}$. 
In the case, as is seen in Eqs.~(\ref{SNRc}) and (\ref{SNR}), 
because the generation rate of the AS photons is enhanced 
by $F\u{c}/2$ within the bandwidth, 
the cavity enhancement of the conversion efficiency 
by $F\u{c}/\pi$ leads to the degradation of the SNR 
by a factor of $2/\pi$ totally, 
compared with the conventional QFC followed by the BPF of $\Delta\u{BPF}=\delta\u{FWHM}$. 
Another candidate to improve the SNR of QFC is the cavity confinement of the signal photon, with the finesse $F\u{s}$ of the cavity. 
In this case, regardless of $\Delta\u{BPF}$, 
only the QFC efficiency is enhanced by $F\u{s}/\pi$, 
whereas the SPDC process generating the AS photons is not enhanced. 
Thus, the SNR is improved 
by $F\u{s}/\pi$ for $\Delta\u{BPF}=\delta\u{FSR}$ 
and $F\u{c}F\u{s}/\pi$ for $\Delta\u{BPF}=\delta\u{FWHM}$. 
The relations among the different configurations of the QFC setups are 
shown in Table~\ref{tbl:SNR}. 
We note that we do not include the confinement of the pump light~\cite{ikuta2021cavity,mann2023low,geus2024low} here, 
because it only enhances the pump light intensity inside the resonator 
and does not change the relationship 
between the SNR and the conversion efficiency. 

Until here, we have considered the QFC 
where the converted and the pump frequencies are close 
with satisfying $\omega\u{s}/2 > \omega\u{p}$, 
and treated the AS photons produced by the pump light as the dominant noise source. 
The similar concept of the SNR improvement by the cavity enhancement 
can be applied to the cases satisfying $\omega\u{p} \gg \omega\u{c}$ 
such as the QFC of single photons at \SI{637}{nm} emitted from an NV center in a diamond.
In the reported experiments~\cite{dreau2018quantum,tchebotareva2019entanglement,stolk2022telecom}, 
the \SI{637}{nm} photon was converted to the \SI{1587}{nm} photon 
using a PPLN waveguide with a \SI{1064}{nm} strong pump light.
In these cases, the dominant noise source contaminating the converted photon
is thought to be SPDC photon pairs 
in the vicinity of \SI{1587}{nm} and \SI{3229}{nm} produced by the pump light. 
We consider introducing the cavity structure around \SI{3229}{nm} to the PPLN, which causes the singly resonant cavity-enhanced SPDC process~\cite{slattery2019background,ikuta2019frequency}.
Since the pump light at \SI{1064}{nm} is the cw light in conventional setups, 
the spectrum around \SI{1587}{nm} exhibits a comb-like structure~\cite{jeronimo2010theory}
despite the absence of an actual cavity around \SI{1587}{nm}. 
The DFG from \SI{637}{nm} to \SI{1587}{nm} and the above singly-resonant SPDC processes are independent. Thus, 
the fine-tuning of the temperature of the PPLN or the pump frequency 
allows the non-resonant region of the SPDC photon pairs 
to be the converted wavelength at \SI{1587}{nm}, without affecting the phase matching of the DFG. 
Consequently, the SNR improvement of the QFC will be possible.
As an example, when the cavity finesse around \SI{3229}{nm} is 45 
with rear and front reflectances are 1 and 0.86, 
the FSR is \SI{5}{\giga Hz}, and $\Delta\u{BPF}=\SI{0.03}{\nano\meter}$, 
the SNR can be enhanced by over tenfold compared to conventional methods.

\section{Conclusion}
In this paper, 
we presented the QFC with cavity enhancement of the converted mode. 
We derived the SNR of the QFC by focusing on the case where the noise source is the AS photons produced by the strong pump light. 
We showed that the cavity structure can significantly improve the SNR 
using feasible parameters compared with the SNR of QFC without the cavity, 
when the photon detection bandwidth is comparable to the FSR. 
We experimentally performed the QFC of a single photon from \SI{780}{nm} to \SI{1540}{nm} based on the PPLN-WR which only confined converted mode. 
While the linewidth of the input photon is much wider than the FWHM of the cavity, we successfully observed the nonclassical photon statistics after the cavity-enhanced QFC. 
By applying the concept of the QFC with the cavity structure for certain frequency modes, we showed the possibility of the SNR improvement of QFCs 
where the input signal photon is confined or the noise source is the SPDC photons. 
We believe the low-noise property of the cavity-enhanced QFCs will be useful in establishing entanglement in the quantum internet, including heterogeneous quantum systems. 

\section*{Funding}
This work was supported by Moonshot R \& D, JST JPMJMS2066, JST JPMJMS226C; FOREST Program, JST JPMJFR222V; R \& D of ICT Priority Technology Project JPMI00316, Asahi Glass Foundation, JSPS JP22J20801, and Program for Leading Graduate Schools: Interactive Materials Science Cadet Program.

\section*{Acknowledgments}
T.Y. and R.I. acknowledge the members of the Quantum Internet Task Force for the comprehensive and interdisciplinary discussions on the quantum internet.

\appendix
\section*{Appendix}
\section{Evaluation of the FSR of the PPLN-WR}
\label{app:A}
\begin{figure}[t]
\centering\includegraphics[width=7cm]{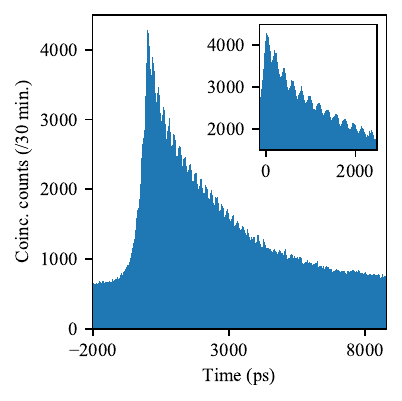}
\caption{The coincidence counts of the SPDC photon-pairs}
\label{fig:spdc}
\end{figure}
We evaluated the FSR of the cavity using the singly-resonant SPDC 
at \SI{1540}{nm} and \SI{1581}{nm} photon pairs pumped by a \SI{780}{nm} laser light. 
Fig.~\ref{fig:spdc} is the result of the observed coincidence counts 
which shows the clear oscillation caused by the cavity structure~\cite{ikuta2019frequency}. 
We performed the Fourier transform for this data, from which the frequency of the oscillation was estimated to be 5.2 $\pm$ \SI{0.1}{\giga\Hz} as the FSR of the cavity.
The estimated value is in good agreement with the results of the resonant structure of the AS photons shown in Figs.~\ref{fig:spectrum} and \ref{fig:spectrum_1522}. 

\section{Reflectance of the PPLN-WR}
\label{app:B}
\begin{figure}[t]
\centering\includegraphics{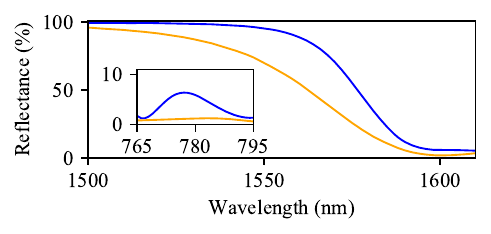}
\caption{The data of the reflectance of the front (blue) and rear (orange) sides of the PPLN-WR. }
\label{fig:reflectivity}
\end{figure}
Fig.~\ref{fig:reflectivity} shows the reflectance of the rear and front sides of the PPLN-WR.
The reflectances are measured by using LN samples loaded in the same batches 
for the coatings. 
As is described in the main text, 
the reflectances are asymmetric to extract the converted photon 
from the rear side efficiently~\cite{pomarico2009waveguide}. 
For the signal light at \SI{780}{nm}, antireflective coatings are applied. 
For the pump light at \SI{1581}{nm}, the PPLN-WR has a slightly resonant structure. 
We utilize the slope of the small transmission peak of the \SI{1581}{nm} light 
for the frequency stabilization of \SI{1540}{nm} light 
to maximize the conversion efficiency. 

\section{The cavity enhancement effect on frequency conversion to \SI{1522}{\nano\meter} light}
\label{app:C}
\begin{figure}[t]
\centering\includegraphics{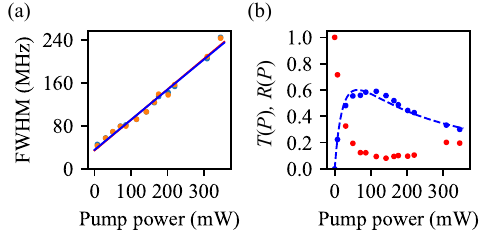}
\caption{(a) Pump power dependencies of observed bandwidths of transmitted~(red) and converted~(blue) light for the \SI{1600}{nm}. 
(b) Pump power dependencies of $T(P) = |t_\mathrm{ss}|^2$~(red) and $R(P) = |r_\mathrm{rs}|^2$ for $\tilde{\Delta} = 0$. 
The blue dashed line was obtained from Eq.~(\ref{ref_prob}) 
using $(\tilde{\alpha}^\U{ex})^{-1}\sim$ \SI{61}{mW} and $\tilde{\gamma}\u{r} = 0.7$.}
\label{fig:conveff_1522}
\end{figure}
We confirmed the cavity enhancement effect of frequency conversion 
to the \SI{1522}{nm} light using the \SI{780}{nm} laser light and the \SI{1600}{nm} pump light. 
Because the front and rear reflectances for \SI{1522}{nm} are 
\SI{99}{\%} and \SI{91}{\%}, respectively, 
a higher enhancement effect on the conversion efficiency is expected.  
Fig.~\ref{fig:conveff_1522}~(a) shows the pump power dependency of the FWHMs of the \SI{780}{nm} signal light and the \SI{1522}{nm} converted light. 
By fitting the experimental results with a function 
$\alpha^\U{ex}_{1522} P + \gamma_\mathrm{all,1522}^\mathrm{ex}$, 
we obtained 
$\alpha^\U{ex}_{1522} = \SI{0.56}{MHz/mW}$ and 
$\gamma_\mathrm{all,1522}^\mathrm{ex} = \SI{34.4}{MHz}$.
The finesse without the pump light is estimated to be 
$F\u{cold,1522}^\U{ex}=\delta\u{FSR}/\gamma_\mathrm{all,1522}^\mathrm{ex} = 151$. 
The pump power dependencies of the internal conversion efficiencies 
are described in Fig.~\ref{fig:conveff_1522}~(b), 
which is in good agreement with Eq.~(\ref{ref_prob}) using
$\alpha^\U{ex}$ and $\gamma_\mathrm{all,1522}^\mathrm{ex}$. 
The maximum conversion efficiency is achieved 
at $(\tilde{\alpha}^\U{ex}_{1522})^{-1}=\gamma\u{all,1522}^\U{ex}/\alpha_{1522} \sim $ \SI{61}{mW}.
The QFC from \SI{780}{nm} to \SI{1522}{nm} 
without the cavity was reported in Refs.~\cite{ikuta2011wide,ikuta2013high-fidelity}. 
Borrowing the reported value of $B^\U{ref}_{1522}=\SI{3.6e-3}{\mbox{/}mW}$ 
and the length $L^\U{ref}_{1522}=\SI{20}{mm}$ of the PPLN waveguide, 
the enhancement factor of the PPLN-WR in our experiment is estimated to be 
$4\tilde{\alpha}\u{ex,1522}/B^\U{ref}_{1522} (L^\U{ref}_{1522}/L)^{2} = 41$. 
The value agrees well with $F\u{cold,1522}^\U{ex}/\pi \sim 48$. 

\begin{figure*}[t]
\centering\includegraphics{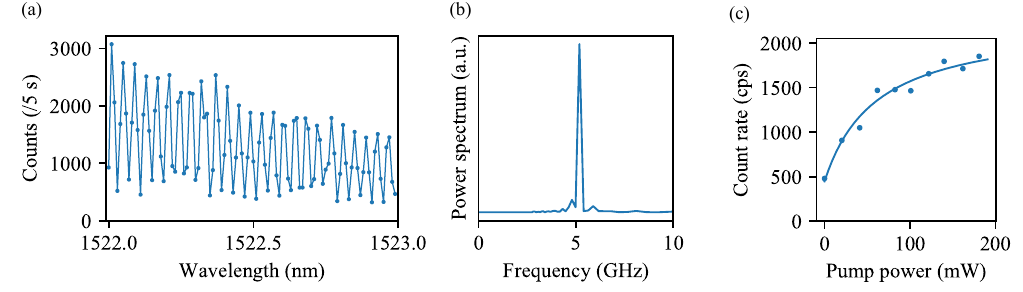}
\caption{(a) Photon count distribution of AS photon around 1522 nm and
(b) its power spectra. (c) The pump power dependency of the number of AS photons. The blue line is calculated using Eq.~(\ref{eq:noise_pump}) and the data of bandwidths of \SI{1522}{nm} converted light.} 
\label{fig:spectrum_1522}
\end{figure*}
We also investigated the properties of the AS photons around \SI{1522}{nm} 
by the measurement similar to Fig.~\ref{fig:spectrum} 
for the AS photons around \SI{1540}{nm}. 
This result of the photon counts in Fig.~\ref{fig:spectrum_1522}~(a) 
clearly shows the oscillation. 
From the Fourier transform on the data depicted in Fig.~\ref{fig:spectrum_1522}~(b), 
the period was found to be $5.2 \pm 0.2$\si{\giga\Hz}, 
which corresponds to the FSR of the cavity.

We measured the pump power dependency of the AS photons 
with $\Delta\u{BPF}=\SI{3.88}{GHz}$. 
The result is shown in Fig.~\ref{fig:spectrum_1522}~(c). 
By fitting the data using Eq.~(\ref{eq:noise_pump}), 
we obtain $\alpha_\mathrm{noise,1522}^\U{ex} = \SI{85}{cps/\milli\watt}$. 
We compared the result with the reported value 
$\alpha\u{noise}^\U{ref} =\SI{80}{cps/\milli\watt}$~\cite{ikuta2013high-fidelity}. 
The normalized proportional coefficients just after the PPLNs were estimated to be 
$\alpha\u{noise,1522}/(L \Delta\u{BPF} T\u{circ}\tilde{\gamma}\u{r})\sim$
\SI{20}{cps\per(\milli\meter\,\giga\Hz\,\milli W)}
and 
$\alpha\u{noise,1522}^\U{ref}/(L^\U{ref}_{1522}\Delta\u{BPF,1522}^{\U{ref}}T\u{circ,1522}^\U{ref})\sim $
\SI{3.2}{cps\per(\milli\meter\,\giga\Hz\,\milli W)}, 
where $\tilde{\gamma}\u{r} = 0.7$$, \Delta\u{BPF,1522}^\U{ref} = \SI{92}{\giga\Hz}$ and $T\u{circ,1522}^\U{ref}=0.014$. 
While the estimated coefficient is slightly larger than we expected, 
the difference is within one order of magnitude.

\section{Pump power dependency of AS photons}
\label{app:ASphoton}
Here, we derive the pump power dependency on the amount of AS photons. 
As written in Section~\ref{sec:theory}, 
we treat the AS photon generation process as the SFG of the phonons. 
We assume the SFG is independent of QFC 
because the broadly generated AS photons and the narrow converted photon 
rarely interfere with each other. 
We also assume the coupling strength between the phonon and the converted modes, 
denoted by $|\Gamma|$, is much smaller than the coupling strength $|\xi|$ 
for QFC as $|\xi|\gg |\Gamma|$ because the SFG is far from the phase matching condition. Under the assumptions, 
the time evolution of the cavity mode is described by
\begin{equation}
\od{\hat{a}\u{c}}{t} = \left( i\Delta\u{c} -\frac{\gamma\u{all}+|\xi|^2}{2}\right)
\hat{a}\u{c} + \Gamma \hat{a}\u{pn,in}.
\end{equation}
$\hat{a}\u{pn,in}$ is the annihilator operator of the input phonon mode.
Using the input-output relation $\hat{a}\u{r,in}+\hat{a}\u{r,out} = \sqrt{\gamma\u{r}}\hat{a}\u{c}$ with $\hat{a}\u{r,in}=0$, 
the mean photon number of the AS photons outside the cavity is described by
\begin{align}\label{eq:intensity_as}
    \braket{\hat{a}^{\dagger}\u{r,out}\hat{a}\u{r,out}} 
    &= \tilde{\gamma\u{r}}
    \expect{\hat{a}\u{pn,in}^\dagger\hat{a}\u{pn,in}} \\
    &= \tilde{\gamma\u{r}}\frac{\tilde{\beta} P}{\frac{1}{4}(1 + \tilde{\alpha}P)^2 + \tilde{\Delta}\u{c}^2}, 
    \label{eq:beta}
\end{align}
where 
$\tilde{\beta} P = |\Gamma|^2 \expect{\hat{a}\u{pn,in}^\dagger\hat{a}\u{pn,in}}/\gamma\u{all}$. 
As a result, the total amount of the AS photons 
within the range of $\delta_\mathrm{FSR}~(\gg \delta_\mathrm{FWHM})$ 
is described by 
\begin{align}
N\u{noise,cav} &= \int^{\delta_\mathrm{FSR}/2}_{-\delta_\mathrm{FSR}/2} 
    \braket{\hat{a}^{\dagger}\u{r,out}\hat{a}\u{r,out}} \U{d}\Delta_\mathrm{c} \nonumber\\
    &\sim \int^{\infty}_{-\infty} 
    \braket{\hat{a}^{\dagger}\u{r,out}\hat{a}\u{r,out}} \U{d}\Delta_\mathrm{c}
    \nonumber \\ 
&= \frac{2 \pi \tilde{r}\u{r}\gamma\u{all} \tilde{\beta} P}{1+\tilde{\alpha}P}. 
\label{eq:ncav}
\end{align}

For QFCs without the cavity, 
the AS photons within the range of $\Delta_\mathrm{BPF}=\delta\u{FSR}$ is described by $N\u{noise,nocav} = \alpha\u{noise} P$ 
in the low pump power regime. In the regime, 
similar to the discussion in Section~\ref{sec:theory}, 
the cavity enhancement of the SFG at the resonant frequency 
with a factor of $F\u{cold}B'/\pi$ leads to 
$\tilde{\beta} = F\u{cold}\alpha\u{noise}/(4\pi\delta\u{FSR})$, 
from the comparison of 
Eq.~(\ref{eq:beta}) with $\tilde{\Delta}\u{c}=0$ and 
$N\u{noise,nocav}/\delta\u{FSR}$. 
Using the result and Eq.~(\ref{eq:ncav}), we obtain
\begin{equation}\label{eq:noise_pump}
N\u{noise,cav} = \frac{\tilde{\gamma}\u{r}\alpha\u{noise}P} {2(1+\tilde{\alpha}P)}. 
\end{equation}
This is bounded from the above by 
$\tilde{\gamma}\u{r}\alpha\u{noise}P/2$
obtained without the effect of the frequency up-conversion of the AS photons 
in the low pump power regime. 
In Section~\ref{sec:theory}, we discuss the SNR 
using the upper bound as $N\u{noise,cav}$ for simplicity.

\bibliography{./aipsamp}

\begin{thebibliography}{40}%
\makeatletter
\providecommand \@ifxundefined [1]{%
 \@ifx{#1\undefined}
}%
\providecommand \@ifnum [1]{%
 \ifnum #1\expandafter \@firstoftwo
 \else \expandafter \@secondoftwo
 \fi
}%
\providecommand \@ifx [1]{%
 \ifx #1\expandafter \@firstoftwo
 \else \expandafter \@secondoftwo
 \fi
}%
\providecommand \natexlab [1]{#1}%
\providecommand \enquote  [1]{``#1''}%
\providecommand \bibnamefont  [1]{#1}%
\providecommand \bibfnamefont [1]{#1}%
\providecommand \citenamefont [1]{#1}%
\providecommand \href@noop [0]{\@secondoftwo}%
\providecommand \href [0]{\begingroup \@sanitize@url \@href}%
\providecommand \@href[1]{\@@startlink{#1}\@@href}%
\providecommand \@@href[1]{\endgroup#1\@@endlink}%
\providecommand \@sanitize@url [0]{\catcode `\\12\catcode `\$12\catcode
  `\&12\catcode `\#12\catcode `\^12\catcode `\_12\catcode `\%12\relax}%
\providecommand \@@startlink[1]{}%
\providecommand \@@endlink[0]{}%
\providecommand \url  [0]{\begingroup\@sanitize@url \@url }%
\providecommand \@url [1]{\endgroup\@href {#1}{\urlprefix }}%
\providecommand \urlprefix  [0]{URL }%
\providecommand \Eprint [0]{\href }%
\providecommand \doibase [0]{http://dx.doi.org/}%
\providecommand \selectlanguage [0]{\@gobble}%
\providecommand \bibinfo  [0]{\@secondoftwo}%
\providecommand \bibfield  [0]{\@secondoftwo}%
\providecommand \translation [1]{[#1]}%
\providecommand \BibitemOpen [0]{}%
\providecommand \bibitemStop [0]{}%
\providecommand \bibitemNoStop [0]{.\EOS\space}%
\providecommand \EOS [0]{\spacefactor3000\relax}%
\providecommand \BibitemShut  [1]{\csname bibitem#1\endcsname}%
\let\auto@bib@innerbib\@empty
\bibitem [{\citenamefont {De~Leon}\ \emph {et~al.}(2021)\citenamefont
  {De~Leon}, \citenamefont {Itoh}, \citenamefont {Kim}, \citenamefont {Mehta},
  \citenamefont {Northup}, \citenamefont {Paik}, \citenamefont {Palmer},
  \citenamefont {Samarth}, \citenamefont {Sangtawesin},\ and\ \citenamefont
  {Steuerman}}]{de2021materials}%
  \BibitemOpen
  \bibfield  {author} {\bibinfo {author} {\bibfnamefont {N.~P.}\ \bibnamefont
  {De~Leon}}, \bibinfo {author} {\bibfnamefont {K.~M.}\ \bibnamefont {Itoh}},
  \bibinfo {author} {\bibfnamefont {D.}~\bibnamefont {Kim}}, \bibinfo {author}
  {\bibfnamefont {K.~K.}\ \bibnamefont {Mehta}}, \bibinfo {author}
  {\bibfnamefont {T.~E.}\ \bibnamefont {Northup}}, \bibinfo {author}
  {\bibfnamefont {H.}~\bibnamefont {Paik}}, \bibinfo {author} {\bibfnamefont
  {B.}~\bibnamefont {Palmer}}, \bibinfo {author} {\bibfnamefont
  {N.}~\bibnamefont {Samarth}}, \bibinfo {author} {\bibfnamefont
  {S.}~\bibnamefont {Sangtawesin}}, \ and\ \bibinfo {author} {\bibfnamefont
  {D.~W.}\ \bibnamefont {Steuerman}},\ }\href@noop {} {\bibfield  {journal}
  {\bibinfo  {journal} {Science}\ }\textbf {\bibinfo {volume} {372}},\ \bibinfo
  {pages} {eabb2823} (\bibinfo {year} {2021})}\BibitemShut {NoStop}%
\bibitem [{\citenamefont {Bluvstein}\ \emph {et~al.}(2024)\citenamefont
  {Bluvstein}, \citenamefont {Evered}, \citenamefont {Geim}, \citenamefont
  {Li}, \citenamefont {Zhou}, \citenamefont {Manovitz}, \citenamefont {Ebadi},
  \citenamefont {Cain}, \citenamefont {Kalinowski}, \citenamefont {Hangleiter}
  \emph {et~al.}}]{bluvstein2024logical}%
  \BibitemOpen
  \bibfield  {author} {\bibinfo {author} {\bibfnamefont {D.}~\bibnamefont
  {Bluvstein}}, \bibinfo {author} {\bibfnamefont {S.~J.}\ \bibnamefont
  {Evered}}, \bibinfo {author} {\bibfnamefont {A.~A.}\ \bibnamefont {Geim}},
  \bibinfo {author} {\bibfnamefont {S.~H.}\ \bibnamefont {Li}}, \bibinfo
  {author} {\bibfnamefont {H.}~\bibnamefont {Zhou}}, \bibinfo {author}
  {\bibfnamefont {T.}~\bibnamefont {Manovitz}}, \bibinfo {author}
  {\bibfnamefont {S.}~\bibnamefont {Ebadi}}, \bibinfo {author} {\bibfnamefont
  {M.}~\bibnamefont {Cain}}, \bibinfo {author} {\bibfnamefont {M.}~\bibnamefont
  {Kalinowski}}, \bibinfo {author} {\bibfnamefont {D.}~\bibnamefont
  {Hangleiter}},  \emph {et~al.},\ }\href@noop {} {\bibfield  {journal}
  {\bibinfo  {journal} {Nature}\ }\textbf {\bibinfo {volume} {626}},\ \bibinfo
  {pages} {58} (\bibinfo {year} {2024})}\BibitemShut {NoStop}%
\bibitem [{\citenamefont {Kimble}(2008)}]{kimble2008quantum}%
  \BibitemOpen
  \bibfield  {author} {\bibinfo {author} {\bibfnamefont {H.~J.}\ \bibnamefont
  {Kimble}},\ }\href@noop {} {\bibfield  {journal} {\bibinfo  {journal}
  {Nature}\ }\textbf {\bibinfo {volume} {453}},\ \bibinfo {pages} {1023}
  (\bibinfo {year} {2008})}\BibitemShut {NoStop}%
\bibitem [{\citenamefont {Wehner}\ \emph {et~al.}(2018)\citenamefont {Wehner},
  \citenamefont {Elkouss},\ and\ \citenamefont {Hanson}}]{wehner2018quantum}%
  \BibitemOpen
  \bibfield  {author} {\bibinfo {author} {\bibfnamefont {S.}~\bibnamefont
  {Wehner}}, \bibinfo {author} {\bibfnamefont {D.}~\bibnamefont {Elkouss}}, \
  and\ \bibinfo {author} {\bibfnamefont {R.}~\bibnamefont {Hanson}},\
  }\href@noop {} {\bibfield  {journal} {\bibinfo  {journal} {Science}\ }\textbf
  {\bibinfo {volume} {362}},\ \bibinfo {pages} {eaam9288} (\bibinfo {year}
  {2018})}\BibitemShut {NoStop}%
\bibitem [{\citenamefont {Wei}\ \emph {et~al.}(2022)\citenamefont {Wei},
  \citenamefont {Jing}, \citenamefont {Zhang}, \citenamefont {Liao},
  \citenamefont {Yuan}, \citenamefont {Fan}, \citenamefont {Lyu}, \citenamefont
  {Zhou}, \citenamefont {Wang}, \citenamefont {Deng} \emph
  {et~al.}}]{wei2022towards}%
  \BibitemOpen
  \bibfield  {author} {\bibinfo {author} {\bibfnamefont {S.-H.}\ \bibnamefont
  {Wei}}, \bibinfo {author} {\bibfnamefont {B.}~\bibnamefont {Jing}}, \bibinfo
  {author} {\bibfnamefont {X.-Y.}\ \bibnamefont {Zhang}}, \bibinfo {author}
  {\bibfnamefont {J.-Y.}\ \bibnamefont {Liao}}, \bibinfo {author}
  {\bibfnamefont {C.-Z.}\ \bibnamefont {Yuan}}, \bibinfo {author}
  {\bibfnamefont {B.-Y.}\ \bibnamefont {Fan}}, \bibinfo {author} {\bibfnamefont
  {C.}~\bibnamefont {Lyu}}, \bibinfo {author} {\bibfnamefont {D.-L.}\
  \bibnamefont {Zhou}}, \bibinfo {author} {\bibfnamefont {Y.}~\bibnamefont
  {Wang}}, \bibinfo {author} {\bibfnamefont {G.-W.}\ \bibnamefont {Deng}},
  \emph {et~al.},\ }\href@noop {} {\bibfield  {journal} {\bibinfo  {journal}
  {Laser \& Photonics Reviews}\ }\textbf {\bibinfo {volume} {16}},\ \bibinfo
  {pages} {2100219} (\bibinfo {year} {2022})}\BibitemShut {NoStop}%
\bibitem [{\citenamefont {Kumar}(1990)}]{kumar1990quantum}%
  \BibitemOpen
  \bibfield  {author} {\bibinfo {author} {\bibfnamefont {P.}~\bibnamefont
  {Kumar}},\ }\href@noop {} {\bibfield  {journal} {\bibinfo  {journal} {Optics
  letters}\ }\textbf {\bibinfo {volume} {15}},\ \bibinfo {pages} {1476}
  (\bibinfo {year} {1990})}\BibitemShut {NoStop}%
\bibitem [{\citenamefont {Ikuta}\ \emph {et~al.}(2011)\citenamefont {Ikuta},
  \citenamefont {Kusaka}, \citenamefont {Kitano}, \citenamefont {Kato},
  \citenamefont {Yamamoto}, \citenamefont {Koashi},\ and\ \citenamefont
  {Imoto}}]{ikuta2011wide}%
  \BibitemOpen
  \bibfield  {author} {\bibinfo {author} {\bibfnamefont {R.}~\bibnamefont
  {Ikuta}}, \bibinfo {author} {\bibfnamefont {Y.}~\bibnamefont {Kusaka}},
  \bibinfo {author} {\bibfnamefont {T.}~\bibnamefont {Kitano}}, \bibinfo
  {author} {\bibfnamefont {H.}~\bibnamefont {Kato}}, \bibinfo {author}
  {\bibfnamefont {T.}~\bibnamefont {Yamamoto}}, \bibinfo {author}
  {\bibfnamefont {M.}~\bibnamefont {Koashi}}, \ and\ \bibinfo {author}
  {\bibfnamefont {N.}~\bibnamefont {Imoto}},\ }\href@noop {} {\bibfield
  {journal} {\bibinfo  {journal} {Nature Communications}\ }\textbf {\bibinfo
  {volume} {2}},\ \bibinfo {pages} {537} (\bibinfo {year} {2011})}\BibitemShut
  {NoStop}%
\bibitem [{\citenamefont {Ikuta}\ \emph {et~al.}(2018)\citenamefont {Ikuta},
  \citenamefont {Kobayashi}, \citenamefont {Kawakami}, \citenamefont {Miki},
  \citenamefont {Yabuno}, \citenamefont {Yamashita}, \citenamefont {Terai},
  \citenamefont {Koashi}, \citenamefont {Mukai}, \citenamefont {Yamamoto} \emph
  {et~al.}}]{ikuta2018polarization}%
  \BibitemOpen
  \bibfield  {author} {\bibinfo {author} {\bibfnamefont {R.}~\bibnamefont
  {Ikuta}}, \bibinfo {author} {\bibfnamefont {T.}~\bibnamefont {Kobayashi}},
  \bibinfo {author} {\bibfnamefont {T.}~\bibnamefont {Kawakami}}, \bibinfo
  {author} {\bibfnamefont {S.}~\bibnamefont {Miki}}, \bibinfo {author}
  {\bibfnamefont {M.}~\bibnamefont {Yabuno}}, \bibinfo {author} {\bibfnamefont
  {T.}~\bibnamefont {Yamashita}}, \bibinfo {author} {\bibfnamefont
  {H.}~\bibnamefont {Terai}}, \bibinfo {author} {\bibfnamefont
  {M.}~\bibnamefont {Koashi}}, \bibinfo {author} {\bibfnamefont
  {T.}~\bibnamefont {Mukai}}, \bibinfo {author} {\bibfnamefont
  {T.}~\bibnamefont {Yamamoto}},  \emph {et~al.},\ }\href@noop {} {\bibfield
  {journal} {\bibinfo  {journal} {Nature Communications}\ }\textbf {\bibinfo
  {volume} {9}},\ \bibinfo {pages} {1997} (\bibinfo {year} {2018})}\BibitemShut
  {NoStop}%
\bibitem [{\citenamefont {van Leent}\ \emph {et~al.}(2020)\citenamefont {van
  Leent}, \citenamefont {Bock}, \citenamefont {Garthoff}, \citenamefont
  {Redeker}, \citenamefont {Zhang}, \citenamefont {Bauer}, \citenamefont
  {Rosenfeld}, \citenamefont {Becher},\ and\ \citenamefont
  {Weinfurter}}]{van2020long}%
  \BibitemOpen
  \bibfield  {author} {\bibinfo {author} {\bibfnamefont {T.}~\bibnamefont {van
  Leent}}, \bibinfo {author} {\bibfnamefont {M.}~\bibnamefont {Bock}}, \bibinfo
  {author} {\bibfnamefont {R.}~\bibnamefont {Garthoff}}, \bibinfo {author}
  {\bibfnamefont {K.}~\bibnamefont {Redeker}}, \bibinfo {author} {\bibfnamefont
  {W.}~\bibnamefont {Zhang}}, \bibinfo {author} {\bibfnamefont
  {T.}~\bibnamefont {Bauer}}, \bibinfo {author} {\bibfnamefont
  {W.}~\bibnamefont {Rosenfeld}}, \bibinfo {author} {\bibfnamefont
  {C.}~\bibnamefont {Becher}}, \ and\ \bibinfo {author} {\bibfnamefont
  {H.}~\bibnamefont {Weinfurter}},\ }\href@noop {} {\bibfield  {journal}
  {\bibinfo  {journal} {Physical Review Letters}\ }\textbf {\bibinfo {volume}
  {124}},\ \bibinfo {pages} {010510} (\bibinfo {year} {2020})}\BibitemShut
  {NoStop}%
\bibitem [{\citenamefont {Yu}\ \emph {et~al.}(2020)\citenamefont {Yu},
  \citenamefont {Ma}, \citenamefont {Luo}, \citenamefont {Jing}, \citenamefont
  {Sun}, \citenamefont {Fang}, \citenamefont {Yang}, \citenamefont {Liu},
  \citenamefont {Zheng}, \citenamefont {Xie} \emph
  {et~al.}}]{yu2020entanglement}%
  \BibitemOpen
  \bibfield  {author} {\bibinfo {author} {\bibfnamefont {Y.}~\bibnamefont
  {Yu}}, \bibinfo {author} {\bibfnamefont {F.}~\bibnamefont {Ma}}, \bibinfo
  {author} {\bibfnamefont {X.-Y.}\ \bibnamefont {Luo}}, \bibinfo {author}
  {\bibfnamefont {B.}~\bibnamefont {Jing}}, \bibinfo {author} {\bibfnamefont
  {P.-F.}\ \bibnamefont {Sun}}, \bibinfo {author} {\bibfnamefont {R.-Z.}\
  \bibnamefont {Fang}}, \bibinfo {author} {\bibfnamefont {C.-W.}\ \bibnamefont
  {Yang}}, \bibinfo {author} {\bibfnamefont {H.}~\bibnamefont {Liu}}, \bibinfo
  {author} {\bibfnamefont {M.-Y.}\ \bibnamefont {Zheng}}, \bibinfo {author}
  {\bibfnamefont {X.-P.}\ \bibnamefont {Xie}},  \emph {et~al.},\ }\href@noop {}
  {\bibfield  {journal} {\bibinfo  {journal} {Nature}\ }\textbf {\bibinfo
  {volume} {578}},\ \bibinfo {pages} {240} (\bibinfo {year}
  {2020})}\BibitemShut {NoStop}%
\bibitem [{\citenamefont {van Leent}\ \emph {et~al.}(2022)\citenamefont {van
  Leent}, \citenamefont {Bock}, \citenamefont {Fertig}, \citenamefont
  {Garthoff}, \citenamefont {Eppelt}, \citenamefont {Zhou}, \citenamefont
  {Malik}, \citenamefont {Seubert}, \citenamefont {Bauer}, \citenamefont
  {Rosenfeld} \emph {et~al.}}]{van2022entangling}%
  \BibitemOpen
  \bibfield  {author} {\bibinfo {author} {\bibfnamefont {T.}~\bibnamefont {van
  Leent}}, \bibinfo {author} {\bibfnamefont {M.}~\bibnamefont {Bock}}, \bibinfo
  {author} {\bibfnamefont {F.}~\bibnamefont {Fertig}}, \bibinfo {author}
  {\bibfnamefont {R.}~\bibnamefont {Garthoff}}, \bibinfo {author}
  {\bibfnamefont {S.}~\bibnamefont {Eppelt}}, \bibinfo {author} {\bibfnamefont
  {Y.}~\bibnamefont {Zhou}}, \bibinfo {author} {\bibfnamefont {P.}~\bibnamefont
  {Malik}}, \bibinfo {author} {\bibfnamefont {M.}~\bibnamefont {Seubert}},
  \bibinfo {author} {\bibfnamefont {T.}~\bibnamefont {Bauer}}, \bibinfo
  {author} {\bibfnamefont {W.}~\bibnamefont {Rosenfeld}},  \emph {et~al.},\
  }\href@noop {} {\bibfield  {journal} {\bibinfo  {journal} {Nature}\ }\textbf
  {\bibinfo {volume} {607}},\ \bibinfo {pages} {69} (\bibinfo {year}
  {2022})}\BibitemShut {NoStop}%
\bibitem [{\citenamefont {Luo}\ \emph {et~al.}(2022)\citenamefont {Luo},
  \citenamefont {Yu}, \citenamefont {Liu}, \citenamefont {Zheng}, \citenamefont
  {Wang}, \citenamefont {Wang}, \citenamefont {Li}, \citenamefont {Jiang},
  \citenamefont {Xie}, \citenamefont {Zhang} \emph
  {et~al.}}]{luo2022postselected}%
  \BibitemOpen
  \bibfield  {author} {\bibinfo {author} {\bibfnamefont {X.-Y.}\ \bibnamefont
  {Luo}}, \bibinfo {author} {\bibfnamefont {Y.}~\bibnamefont {Yu}}, \bibinfo
  {author} {\bibfnamefont {J.-L.}\ \bibnamefont {Liu}}, \bibinfo {author}
  {\bibfnamefont {M.-Y.}\ \bibnamefont {Zheng}}, \bibinfo {author}
  {\bibfnamefont {C.-Y.}\ \bibnamefont {Wang}}, \bibinfo {author}
  {\bibfnamefont {B.}~\bibnamefont {Wang}}, \bibinfo {author} {\bibfnamefont
  {J.}~\bibnamefont {Li}}, \bibinfo {author} {\bibfnamefont {X.}~\bibnamefont
  {Jiang}}, \bibinfo {author} {\bibfnamefont {X.-P.}\ \bibnamefont {Xie}},
  \bibinfo {author} {\bibfnamefont {Q.}~\bibnamefont {Zhang}},  \emph
  {et~al.},\ }\href@noop {} {\bibfield  {journal} {\bibinfo  {journal}
  {Physical Review Letters}\ }\textbf {\bibinfo {volume} {129}},\ \bibinfo
  {pages} {050503} (\bibinfo {year} {2022})}\BibitemShut {NoStop}%
\bibitem [{\citenamefont {Liu}\ \emph {et~al.}(2024)\citenamefont {Liu},
  \citenamefont {Luo}, \citenamefont {Yu}, \citenamefont {Wang}, \citenamefont
  {Wang}, \citenamefont {Hu}, \citenamefont {Li}, \citenamefont {Zheng},
  \citenamefont {Yao}, \citenamefont {Yan} \emph {et~al.}}]{liu2024creation}%
  \BibitemOpen
  \bibfield  {author} {\bibinfo {author} {\bibfnamefont {J.-L.}\ \bibnamefont
  {Liu}}, \bibinfo {author} {\bibfnamefont {X.-Y.}\ \bibnamefont {Luo}},
  \bibinfo {author} {\bibfnamefont {Y.}~\bibnamefont {Yu}}, \bibinfo {author}
  {\bibfnamefont {C.-Y.}\ \bibnamefont {Wang}}, \bibinfo {author}
  {\bibfnamefont {B.}~\bibnamefont {Wang}}, \bibinfo {author} {\bibfnamefont
  {Y.}~\bibnamefont {Hu}}, \bibinfo {author} {\bibfnamefont {J.}~\bibnamefont
  {Li}}, \bibinfo {author} {\bibfnamefont {M.-Y.}\ \bibnamefont {Zheng}},
  \bibinfo {author} {\bibfnamefont {B.}~\bibnamefont {Yao}}, \bibinfo {author}
  {\bibfnamefont {Z.}~\bibnamefont {Yan}},  \emph {et~al.},\ }\href@noop {}
  {\bibfield  {journal} {\bibinfo  {journal} {Nature}\ }\textbf {\bibinfo
  {volume} {629}},\ \bibinfo {pages} {579} (\bibinfo {year}
  {2024})}\BibitemShut {NoStop}%
\bibitem [{\citenamefont {Zhou}\ \emph {et~al.}(2024)\citenamefont {Zhou},
  \citenamefont {Malik}, \citenamefont {Fertig}, \citenamefont {Bock},
  \citenamefont {Bauer}, \citenamefont {van Leent}, \citenamefont {Zhang},
  \citenamefont {Becher},\ and\ \citenamefont {Weinfurter}}]{zhou2024long}%
  \BibitemOpen
  \bibfield  {author} {\bibinfo {author} {\bibfnamefont {Y.}~\bibnamefont
  {Zhou}}, \bibinfo {author} {\bibfnamefont {P.}~\bibnamefont {Malik}},
  \bibinfo {author} {\bibfnamefont {F.}~\bibnamefont {Fertig}}, \bibinfo
  {author} {\bibfnamefont {M.}~\bibnamefont {Bock}}, \bibinfo {author}
  {\bibfnamefont {T.}~\bibnamefont {Bauer}}, \bibinfo {author} {\bibfnamefont
  {T.}~\bibnamefont {van Leent}}, \bibinfo {author} {\bibfnamefont
  {W.}~\bibnamefont {Zhang}}, \bibinfo {author} {\bibfnamefont
  {C.}~\bibnamefont {Becher}}, \ and\ \bibinfo {author} {\bibfnamefont
  {H.}~\bibnamefont {Weinfurter}},\ }\href@noop {} {\bibfield  {journal}
  {\bibinfo  {journal} {PRX Quantum}\ }\textbf {\bibinfo {volume} {5}},\
  \bibinfo {pages} {020307} (\bibinfo {year} {2024})}\BibitemShut {NoStop}%
\bibitem [{\citenamefont {Bock}\ \emph {et~al.}(2018)\citenamefont {Bock},
  \citenamefont {Eich}, \citenamefont {Kucera}, \citenamefont {Kreis},
  \citenamefont {Lenhard}, \citenamefont {Becher},\ and\ \citenamefont
  {Eschner}}]{bock2018high}%
  \BibitemOpen
  \bibfield  {author} {\bibinfo {author} {\bibfnamefont {M.}~\bibnamefont
  {Bock}}, \bibinfo {author} {\bibfnamefont {P.}~\bibnamefont {Eich}}, \bibinfo
  {author} {\bibfnamefont {S.}~\bibnamefont {Kucera}}, \bibinfo {author}
  {\bibfnamefont {M.}~\bibnamefont {Kreis}}, \bibinfo {author} {\bibfnamefont
  {A.}~\bibnamefont {Lenhard}}, \bibinfo {author} {\bibfnamefont
  {C.}~\bibnamefont {Becher}}, \ and\ \bibinfo {author} {\bibfnamefont
  {J.}~\bibnamefont {Eschner}},\ }\href@noop {} {\bibfield  {journal} {\bibinfo
   {journal} {Nature Communications}\ }\textbf {\bibinfo {volume} {9}},\
  \bibinfo {pages} {1998} (\bibinfo {year} {2018})}\BibitemShut {NoStop}%
\bibitem [{\citenamefont {Krutyanskiy}\ \emph {et~al.}(2019)\citenamefont
  {Krutyanskiy}, \citenamefont {Meraner}, \citenamefont {Schupp}, \citenamefont
  {Krcmarsky}, \citenamefont {Hainzer},\ and\ \citenamefont
  {Lanyon}}]{krutyanskiy2019light}%
  \BibitemOpen
  \bibfield  {author} {\bibinfo {author} {\bibfnamefont {V.}~\bibnamefont
  {Krutyanskiy}}, \bibinfo {author} {\bibfnamefont {M.}~\bibnamefont
  {Meraner}}, \bibinfo {author} {\bibfnamefont {J.}~\bibnamefont {Schupp}},
  \bibinfo {author} {\bibfnamefont {V.}~\bibnamefont {Krcmarsky}}, \bibinfo
  {author} {\bibfnamefont {H.}~\bibnamefont {Hainzer}}, \ and\ \bibinfo
  {author} {\bibfnamefont {B.~P.}\ \bibnamefont {Lanyon}},\ }\href@noop {}
  {\bibfield  {journal} {\bibinfo  {journal} {npj Quantum Information}\
  }\textbf {\bibinfo {volume} {5}},\ \bibinfo {pages} {72} (\bibinfo {year}
  {2019})}\BibitemShut {NoStop}%
\bibitem [{\citenamefont {Tchebotareva}\ \emph {et~al.}(2019)\citenamefont
  {Tchebotareva}, \citenamefont {Hermans}, \citenamefont {Humphreys},
  \citenamefont {Voigt}, \citenamefont {Harmsma}, \citenamefont {Cheng},
  \citenamefont {Verlaan}, \citenamefont {Dijkhuizen}, \citenamefont {De~Jong},
  \citenamefont {Dr{\'e}au} \emph {et~al.}}]{tchebotareva2019entanglement}%
  \BibitemOpen
  \bibfield  {author} {\bibinfo {author} {\bibfnamefont {A.}~\bibnamefont
  {Tchebotareva}}, \bibinfo {author} {\bibfnamefont {S.~L.}\ \bibnamefont
  {Hermans}}, \bibinfo {author} {\bibfnamefont {P.~C.}\ \bibnamefont
  {Humphreys}}, \bibinfo {author} {\bibfnamefont {D.}~\bibnamefont {Voigt}},
  \bibinfo {author} {\bibfnamefont {P.~J.}\ \bibnamefont {Harmsma}}, \bibinfo
  {author} {\bibfnamefont {L.~K.}\ \bibnamefont {Cheng}}, \bibinfo {author}
  {\bibfnamefont {A.~L.}\ \bibnamefont {Verlaan}}, \bibinfo {author}
  {\bibfnamefont {N.}~\bibnamefont {Dijkhuizen}}, \bibinfo {author}
  {\bibfnamefont {W.}~\bibnamefont {De~Jong}}, \bibinfo {author} {\bibfnamefont
  {A.}~\bibnamefont {Dr{\'e}au}},  \emph {et~al.},\ }\href@noop {} {\bibfield
  {journal} {\bibinfo  {journal} {Physical Review Letters}\ }\textbf {\bibinfo
  {volume} {123}},\ \bibinfo {pages} {063601} (\bibinfo {year}
  {2019})}\BibitemShut {NoStop}%
\bibitem [{\citenamefont {Bersin}\ \emph {et~al.}(2024)\citenamefont {Bersin},
  \citenamefont {Sutula}, \citenamefont {Huan}, \citenamefont {Suleymanzade},
  \citenamefont {Assumpcao}, \citenamefont {Wei}, \citenamefont {Stas},
  \citenamefont {Knaut}, \citenamefont {Knall}, \citenamefont {Langrock} \emph
  {et~al.}}]{bersin2024telecom}%
  \BibitemOpen
  \bibfield  {author} {\bibinfo {author} {\bibfnamefont {E.}~\bibnamefont
  {Bersin}}, \bibinfo {author} {\bibfnamefont {M.}~\bibnamefont {Sutula}},
  \bibinfo {author} {\bibfnamefont {Y.~Q.}\ \bibnamefont {Huan}}, \bibinfo
  {author} {\bibfnamefont {A.}~\bibnamefont {Suleymanzade}}, \bibinfo {author}
  {\bibfnamefont {D.~R.}\ \bibnamefont {Assumpcao}}, \bibinfo {author}
  {\bibfnamefont {Y.-C.}\ \bibnamefont {Wei}}, \bibinfo {author} {\bibfnamefont
  {P.-J.}\ \bibnamefont {Stas}}, \bibinfo {author} {\bibfnamefont {C.~M.}\
  \bibnamefont {Knaut}}, \bibinfo {author} {\bibfnamefont {E.~N.}\ \bibnamefont
  {Knall}}, \bibinfo {author} {\bibfnamefont {C.}~\bibnamefont {Langrock}},
  \emph {et~al.},\ }\href@noop {} {\bibfield  {journal} {\bibinfo  {journal}
  {PRX Quantum}\ }\textbf {\bibinfo {volume} {5}},\ \bibinfo {pages} {010303}
  (\bibinfo {year} {2024})}\BibitemShut {NoStop}%
\bibitem [{\citenamefont {Knaut}\ \emph {et~al.}(2024)\citenamefont {Knaut},
  \citenamefont {Suleymanzade}, \citenamefont {Wei}, \citenamefont {Assumpcao},
  \citenamefont {Stas}, \citenamefont {Huan}, \citenamefont {Machielse},
  \citenamefont {Knall}, \citenamefont {Sutula}, \citenamefont {Baranes} \emph
  {et~al.}}]{knaut2024entanglement}%
  \BibitemOpen
  \bibfield  {author} {\bibinfo {author} {\bibfnamefont {C.}~\bibnamefont
  {Knaut}}, \bibinfo {author} {\bibfnamefont {A.}~\bibnamefont {Suleymanzade}},
  \bibinfo {author} {\bibfnamefont {Y.-C.}\ \bibnamefont {Wei}}, \bibinfo
  {author} {\bibfnamefont {D.}~\bibnamefont {Assumpcao}}, \bibinfo {author}
  {\bibfnamefont {P.-J.}\ \bibnamefont {Stas}}, \bibinfo {author}
  {\bibfnamefont {Y.}~\bibnamefont {Huan}}, \bibinfo {author} {\bibfnamefont
  {B.}~\bibnamefont {Machielse}}, \bibinfo {author} {\bibfnamefont
  {E.}~\bibnamefont {Knall}}, \bibinfo {author} {\bibfnamefont
  {M.}~\bibnamefont {Sutula}}, \bibinfo {author} {\bibfnamefont
  {G.}~\bibnamefont {Baranes}},  \emph {et~al.},\ }\href@noop {} {\bibfield
  {journal} {\bibinfo  {journal} {Nature}\ }\textbf {\bibinfo {volume} {629}},\
  \bibinfo {pages} {573} (\bibinfo {year} {2024})}\BibitemShut {NoStop}%
\bibitem [{\citenamefont {Pelc}\ \emph {et~al.}(2011)\citenamefont {Pelc},
  \citenamefont {Ma}, \citenamefont {Phillips}, \citenamefont {Zhang},
  \citenamefont {Langrock}, \citenamefont {Slattery}, \citenamefont {Tang},\
  and\ \citenamefont {Fejer}}]{pelc2011long}%
  \BibitemOpen
  \bibfield  {author} {\bibinfo {author} {\bibfnamefont {J.~S.}\ \bibnamefont
  {Pelc}}, \bibinfo {author} {\bibfnamefont {L.}~\bibnamefont {Ma}}, \bibinfo
  {author} {\bibfnamefont {C.}~\bibnamefont {Phillips}}, \bibinfo {author}
  {\bibfnamefont {Q.}~\bibnamefont {Zhang}}, \bibinfo {author} {\bibfnamefont
  {C.}~\bibnamefont {Langrock}}, \bibinfo {author} {\bibfnamefont
  {O.}~\bibnamefont {Slattery}}, \bibinfo {author} {\bibfnamefont
  {X.}~\bibnamefont {Tang}}, \ and\ \bibinfo {author} {\bibfnamefont {M.~M.}\
  \bibnamefont {Fejer}},\ }\href@noop {} {\bibfield  {journal} {\bibinfo
  {journal} {Optics Express}\ }\textbf {\bibinfo {volume} {19}},\ \bibinfo
  {pages} {21445} (\bibinfo {year} {2011})}\BibitemShut {NoStop}%
\bibitem [{\citenamefont {Ikuta}\ \emph {et~al.}(2014)\citenamefont {Ikuta},
  \citenamefont {Kobayashi}, \citenamefont {Yasui}, \citenamefont {Miki},
  \citenamefont {Yamashita}, \citenamefont {Terai}, \citenamefont {Fujiwara},
  \citenamefont {Yamamoto}, \citenamefont {Koashi}, \citenamefont {Sasaki}
  \emph {et~al.}}]{ikuta2014frequency}%
  \BibitemOpen
  \bibfield  {author} {\bibinfo {author} {\bibfnamefont {R.}~\bibnamefont
  {Ikuta}}, \bibinfo {author} {\bibfnamefont {T.}~\bibnamefont {Kobayashi}},
  \bibinfo {author} {\bibfnamefont {S.}~\bibnamefont {Yasui}}, \bibinfo
  {author} {\bibfnamefont {S.}~\bibnamefont {Miki}}, \bibinfo {author}
  {\bibfnamefont {T.}~\bibnamefont {Yamashita}}, \bibinfo {author}
  {\bibfnamefont {H.}~\bibnamefont {Terai}}, \bibinfo {author} {\bibfnamefont
  {M.}~\bibnamefont {Fujiwara}}, \bibinfo {author} {\bibfnamefont
  {T.}~\bibnamefont {Yamamoto}}, \bibinfo {author} {\bibfnamefont
  {M.}~\bibnamefont {Koashi}}, \bibinfo {author} {\bibfnamefont
  {M.}~\bibnamefont {Sasaki}},  \emph {et~al.},\ }\href@noop {} {\bibfield
  {journal} {\bibinfo  {journal} {Optics express}\ }\textbf {\bibinfo {volume}
  {22}},\ \bibinfo {pages} {11205} (\bibinfo {year} {2014})}\BibitemShut
  {NoStop}%
\bibitem [{\citenamefont {Maring}\ \emph {et~al.}(2018)\citenamefont {Maring},
  \citenamefont {Lago-Rivera}, \citenamefont {Lenhard}, \citenamefont
  {Heinze},\ and\ \citenamefont {de~Riedmatten}}]{maring2018quantum}%
  \BibitemOpen
  \bibfield  {author} {\bibinfo {author} {\bibfnamefont {N.}~\bibnamefont
  {Maring}}, \bibinfo {author} {\bibfnamefont {D.}~\bibnamefont {Lago-Rivera}},
  \bibinfo {author} {\bibfnamefont {A.}~\bibnamefont {Lenhard}}, \bibinfo
  {author} {\bibfnamefont {G.}~\bibnamefont {Heinze}}, \ and\ \bibinfo {author}
  {\bibfnamefont {H.}~\bibnamefont {de~Riedmatten}},\ }\href@noop {} {\bibfield
   {journal} {\bibinfo  {journal} {Optica}\ }\textbf {\bibinfo {volume} {5}},\
  \bibinfo {pages} {507} (\bibinfo {year} {2018})}\BibitemShut {NoStop}%
\bibitem [{\citenamefont {Strassmann}\ \emph {et~al.}(2019)\citenamefont
  {Strassmann}, \citenamefont {Martin}, \citenamefont {Gisin},\ and\
  \citenamefont {Afzelius}}]{Strassmann:19}%
  \BibitemOpen
  \bibfield  {author} {\bibinfo {author} {\bibfnamefont {P.~C.}\ \bibnamefont
  {Strassmann}}, \bibinfo {author} {\bibfnamefont {A.}~\bibnamefont {Martin}},
  \bibinfo {author} {\bibfnamefont {N.}~\bibnamefont {Gisin}}, \ and\ \bibinfo
  {author} {\bibfnamefont {M.}~\bibnamefont {Afzelius}},\ }\href {\doibase
  10.1364/OE.27.014298} {\bibfield  {journal} {\bibinfo  {journal} {Opt.
  Express}\ }\textbf {\bibinfo {volume} {27}},\ \bibinfo {pages} {14298}
  (\bibinfo {year} {2019})}\BibitemShut {NoStop}%
\bibitem [{\citenamefont {Walker}\ \emph {et~al.}(2018)\citenamefont {Walker},
  \citenamefont {Miyanishi}, \citenamefont {Ikuta}, \citenamefont {Takahashi},
  \citenamefont {Vartabi~Kashanian}, \citenamefont {Tsujimoto}, \citenamefont
  {Hayasaka}, \citenamefont {Yamamoto}, \citenamefont {Imoto},\ and\
  \citenamefont {Keller}}]{walker2018long}%
  \BibitemOpen
  \bibfield  {author} {\bibinfo {author} {\bibfnamefont {T.}~\bibnamefont
  {Walker}}, \bibinfo {author} {\bibfnamefont {K.}~\bibnamefont {Miyanishi}},
  \bibinfo {author} {\bibfnamefont {R.}~\bibnamefont {Ikuta}}, \bibinfo
  {author} {\bibfnamefont {H.}~\bibnamefont {Takahashi}}, \bibinfo {author}
  {\bibfnamefont {S.}~\bibnamefont {Vartabi~Kashanian}}, \bibinfo {author}
  {\bibfnamefont {Y.}~\bibnamefont {Tsujimoto}}, \bibinfo {author}
  {\bibfnamefont {K.}~\bibnamefont {Hayasaka}}, \bibinfo {author}
  {\bibfnamefont {T.}~\bibnamefont {Yamamoto}}, \bibinfo {author}
  {\bibfnamefont {N.}~\bibnamefont {Imoto}}, \ and\ \bibinfo {author}
  {\bibfnamefont {M.}~\bibnamefont {Keller}},\ }\href@noop {} {\bibfield
  {journal} {\bibinfo  {journal} {Physical review letters}\ }\textbf {\bibinfo
  {volume} {120}},\ \bibinfo {pages} {203601} (\bibinfo {year}
  {2018})}\BibitemShut {NoStop}%
\bibitem [{\citenamefont {Stolk}\ \emph {et~al.}(2022)\citenamefont {Stolk},
  \citenamefont {van~der Enden}, \citenamefont {Roehsner}, \citenamefont
  {Teepe}, \citenamefont {Faes}, \citenamefont {Bradley}, \citenamefont
  {Cadot}, \citenamefont {van Rantwijk}, \citenamefont {Te~Raa}, \citenamefont
  {Hagen} \emph {et~al.}}]{stolk2022telecom}%
  \BibitemOpen
  \bibfield  {author} {\bibinfo {author} {\bibfnamefont {A.}~\bibnamefont
  {Stolk}}, \bibinfo {author} {\bibfnamefont {K.~L.}\ \bibnamefont {van~der
  Enden}}, \bibinfo {author} {\bibfnamefont {M.-C.}\ \bibnamefont {Roehsner}},
  \bibinfo {author} {\bibfnamefont {A.}~\bibnamefont {Teepe}}, \bibinfo
  {author} {\bibfnamefont {S.~O.}\ \bibnamefont {Faes}}, \bibinfo {author}
  {\bibfnamefont {C.}~\bibnamefont {Bradley}}, \bibinfo {author} {\bibfnamefont
  {S.}~\bibnamefont {Cadot}}, \bibinfo {author} {\bibfnamefont
  {J.}~\bibnamefont {van Rantwijk}}, \bibinfo {author} {\bibfnamefont
  {I.}~\bibnamefont {Te~Raa}}, \bibinfo {author} {\bibfnamefont
  {R.}~\bibnamefont {Hagen}},  \emph {et~al.},\ }\href@noop {} {\bibfield
  {journal} {\bibinfo  {journal} {PRX Quantum}\ }\textbf {\bibinfo {volume}
  {3}},\ \bibinfo {pages} {020359} (\bibinfo {year} {2022})}\BibitemShut
  {NoStop}%
\bibitem [{\citenamefont {Ikuta}\ \emph {et~al.}(2022)\citenamefont {Ikuta},
  \citenamefont {Yokota}, \citenamefont {Kobayashi}, \citenamefont {Imoto},\
  and\ \citenamefont {Yamamoto}}]{ikuta2022optical}%
  \BibitemOpen
  \bibfield  {author} {\bibinfo {author} {\bibfnamefont {R.}~\bibnamefont
  {Ikuta}}, \bibinfo {author} {\bibfnamefont {M.}~\bibnamefont {Yokota}},
  \bibinfo {author} {\bibfnamefont {T.}~\bibnamefont {Kobayashi}}, \bibinfo
  {author} {\bibfnamefont {N.}~\bibnamefont {Imoto}}, \ and\ \bibinfo {author}
  {\bibfnamefont {T.}~\bibnamefont {Yamamoto}},\ }\href@noop {} {\bibfield
  {journal} {\bibinfo  {journal} {Physical Review Applied}\ }\textbf {\bibinfo
  {volume} {17}},\ \bibinfo {pages} {034012} (\bibinfo {year}
  {2022})}\BibitemShut {NoStop}%
\bibitem [{\citenamefont {Berger}(1997)}]{berger1997second}%
  \BibitemOpen
  \bibfield  {author} {\bibinfo {author} {\bibfnamefont {V.}~\bibnamefont
  {Berger}},\ }\href@noop {} {\bibfield  {journal} {\bibinfo  {journal} {JOSA
  B}\ }\textbf {\bibinfo {volume} {14}},\ \bibinfo {pages} {1351} (\bibinfo
  {year} {1997})}\BibitemShut {NoStop}%
\bibitem [{\citenamefont {Liscidini}\ and\ \citenamefont
  {Claudio~Andreani}(2006)}]{liscidini2006second}%
  \BibitemOpen
  \bibfield  {author} {\bibinfo {author} {\bibfnamefont {M.}~\bibnamefont
  {Liscidini}}\ and\ \bibinfo {author} {\bibfnamefont {L.}~\bibnamefont
  {Claudio~Andreani}},\ }\href@noop {} {\bibfield  {journal} {\bibinfo
  {journal} {Physical Review E}\ }\textbf {\bibinfo {volume} {73}},\ \bibinfo
  {pages} {016613} (\bibinfo {year} {2006})}\BibitemShut {NoStop}%
\bibitem [{\citenamefont {Murakami}\ \emph {et~al.}(2023)\citenamefont
  {Murakami}, \citenamefont {Fujimoto}, \citenamefont {Kobayashi},
  \citenamefont {Ikuta}, \citenamefont {Inoue}, \citenamefont {Umeki},
  \citenamefont {Miki}, \citenamefont {China}, \citenamefont {Terai},
  \citenamefont {Kasahara} \emph {et~al.}}]{murakami2023quantum}%
  \BibitemOpen
  \bibfield  {author} {\bibinfo {author} {\bibfnamefont {S.}~\bibnamefont
  {Murakami}}, \bibinfo {author} {\bibfnamefont {R.}~\bibnamefont {Fujimoto}},
  \bibinfo {author} {\bibfnamefont {T.}~\bibnamefont {Kobayashi}}, \bibinfo
  {author} {\bibfnamefont {R.}~\bibnamefont {Ikuta}}, \bibinfo {author}
  {\bibfnamefont {A.}~\bibnamefont {Inoue}}, \bibinfo {author} {\bibfnamefont
  {T.}~\bibnamefont {Umeki}}, \bibinfo {author} {\bibfnamefont
  {S.}~\bibnamefont {Miki}}, \bibinfo {author} {\bibfnamefont {F.}~\bibnamefont
  {China}}, \bibinfo {author} {\bibfnamefont {H.}~\bibnamefont {Terai}},
  \bibinfo {author} {\bibfnamefont {R.}~\bibnamefont {Kasahara}},  \emph
  {et~al.},\ }\href@noop {} {\bibfield  {journal} {\bibinfo  {journal} {Optics
  Express}\ }\textbf {\bibinfo {volume} {31}},\ \bibinfo {pages} {29271}
  (\bibinfo {year} {2023})}\BibitemShut {NoStop}%
\bibitem [{\citenamefont {Albrecht}\ \emph {et~al.}(2014)\citenamefont
  {Albrecht}, \citenamefont {Farrera}, \citenamefont {Fernandez-Gonzalvo},
  \citenamefont {Cristiani},\ and\ \citenamefont
  {De~Riedmatten}}]{albrecht2014waveguide}%
  \BibitemOpen
  \bibfield  {author} {\bibinfo {author} {\bibfnamefont {B.}~\bibnamefont
  {Albrecht}}, \bibinfo {author} {\bibfnamefont {P.}~\bibnamefont {Farrera}},
  \bibinfo {author} {\bibfnamefont {X.}~\bibnamefont {Fernandez-Gonzalvo}},
  \bibinfo {author} {\bibfnamefont {M.}~\bibnamefont {Cristiani}}, \ and\
  \bibinfo {author} {\bibfnamefont {H.}~\bibnamefont {De~Riedmatten}},\
  }\href@noop {} {\bibfield  {journal} {\bibinfo  {journal} {Nature
  Communications}\ }\textbf {\bibinfo {volume} {5}},\ \bibinfo {pages} {3376}
  (\bibinfo {year} {2014})}\BibitemShut {NoStop}%
\bibitem [{\citenamefont {Ikuta}\ \emph {et~al.}(2016)\citenamefont {Ikuta},
  \citenamefont {Kobayashi}, \citenamefont {Matsuki}, \citenamefont {Miki},
  \citenamefont {Yamashita}, \citenamefont {Terai}, \citenamefont {Yamamoto},
  \citenamefont {Koashi}, \citenamefont {Mukai},\ and\ \citenamefont
  {Imoto}}]{ikuta2016}%
  \BibitemOpen
  \bibfield  {author} {\bibinfo {author} {\bibfnamefont {R.}~\bibnamefont
  {Ikuta}}, \bibinfo {author} {\bibfnamefont {T.}~\bibnamefont {Kobayashi}},
  \bibinfo {author} {\bibfnamefont {K.}~\bibnamefont {Matsuki}}, \bibinfo
  {author} {\bibfnamefont {S.}~\bibnamefont {Miki}}, \bibinfo {author}
  {\bibfnamefont {T.}~\bibnamefont {Yamashita}}, \bibinfo {author}
  {\bibfnamefont {H.}~\bibnamefont {Terai}}, \bibinfo {author} {\bibfnamefont
  {T.}~\bibnamefont {Yamamoto}}, \bibinfo {author} {\bibfnamefont
  {M.}~\bibnamefont {Koashi}}, \bibinfo {author} {\bibfnamefont
  {T.}~\bibnamefont {Mukai}}, \ and\ \bibinfo {author} {\bibfnamefont
  {N.}~\bibnamefont {Imoto}},\ }\href {\doibase 10.1364/OPTICA.3.001279}
  {\bibfield  {journal} {\bibinfo  {journal} {Optica}\ }\textbf {\bibinfo
  {volume} {3}},\ \bibinfo {pages} {1279} (\bibinfo {year} {2016})}\BibitemShut
  {NoStop}%
\bibitem [{\citenamefont {Ikuta}\ \emph {et~al.}(2021)\citenamefont {Ikuta},
  \citenamefont {Kobayashi}, \citenamefont {Yamazaki}, \citenamefont {Imoto},\
  and\ \citenamefont {Yamamoto}}]{ikuta2021cavity}%
  \BibitemOpen
  \bibfield  {author} {\bibinfo {author} {\bibfnamefont {R.}~\bibnamefont
  {Ikuta}}, \bibinfo {author} {\bibfnamefont {T.}~\bibnamefont {Kobayashi}},
  \bibinfo {author} {\bibfnamefont {T.}~\bibnamefont {Yamazaki}}, \bibinfo
  {author} {\bibfnamefont {N.}~\bibnamefont {Imoto}}, \ and\ \bibinfo {author}
  {\bibfnamefont {T.}~\bibnamefont {Yamamoto}},\ }\href@noop {} {\bibfield
  {journal} {\bibinfo  {journal} {Physical Review A}\ }\textbf {\bibinfo
  {volume} {103}},\ \bibinfo {pages} {033709} (\bibinfo {year}
  {2021})}\BibitemShut {NoStop}%
\bibitem [{\citenamefont {Mann}\ \emph {et~al.}(2023)\citenamefont {Mann},
  \citenamefont {Chrzanowski}, \citenamefont {Gewers}, \citenamefont {Placke},\
  and\ \citenamefont {Ramelow}}]{mann2023low}%
  \BibitemOpen
  \bibfield  {author} {\bibinfo {author} {\bibfnamefont {F.}~\bibnamefont
  {Mann}}, \bibinfo {author} {\bibfnamefont {H.~M.}\ \bibnamefont
  {Chrzanowski}}, \bibinfo {author} {\bibfnamefont {F.}~\bibnamefont {Gewers}},
  \bibinfo {author} {\bibfnamefont {M.}~\bibnamefont {Placke}}, \ and\ \bibinfo
  {author} {\bibfnamefont {S.}~\bibnamefont {Ramelow}},\ }\href@noop {}
  {\bibfield  {journal} {\bibinfo  {journal} {Physical Review Applied}\
  }\textbf {\bibinfo {volume} {20}},\ \bibinfo {pages} {054010} (\bibinfo
  {year} {2023})}\BibitemShut {NoStop}%
\bibitem [{\citenamefont {Geus}\ \emph {et~al.}(2024)\citenamefont {Geus},
  \citenamefont {Elsen}, \citenamefont {Nyga}, \citenamefont {Stolk},
  \citenamefont {van~der Enden}, \citenamefont {van Zwet}, \citenamefont
  {Haefner}, \citenamefont {Hanson},\ and\ \citenamefont
  {Jungbluth}}]{geus2024low}%
  \BibitemOpen
  \bibfield  {author} {\bibinfo {author} {\bibfnamefont {J.~F.}\ \bibnamefont
  {Geus}}, \bibinfo {author} {\bibfnamefont {F.}~\bibnamefont {Elsen}},
  \bibinfo {author} {\bibfnamefont {S.}~\bibnamefont {Nyga}}, \bibinfo {author}
  {\bibfnamefont {A.~J.}\ \bibnamefont {Stolk}}, \bibinfo {author}
  {\bibfnamefont {K.~L.}\ \bibnamefont {van~der Enden}}, \bibinfo {author}
  {\bibfnamefont {E.~J.}\ \bibnamefont {van Zwet}}, \bibinfo {author}
  {\bibfnamefont {C.}~\bibnamefont {Haefner}}, \bibinfo {author} {\bibfnamefont
  {R.}~\bibnamefont {Hanson}}, \ and\ \bibinfo {author} {\bibfnamefont
  {B.}~\bibnamefont {Jungbluth}},\ }\href@noop {} {\bibfield  {journal}
  {\bibinfo  {journal} {Optica Quantum}\ }\textbf {\bibinfo {volume} {2}},\
  \bibinfo {pages} {189} (\bibinfo {year} {2024})}\BibitemShut {NoStop}%
\bibitem [{\citenamefont {Dr{\'e}au}\ \emph {et~al.}(2018)\citenamefont
  {Dr{\'e}au}, \citenamefont {Tchebotareva}, \citenamefont {Mahdaoui},
  \citenamefont {Bonato},\ and\ \citenamefont {Hanson}}]{dreau2018quantum}%
  \BibitemOpen
  \bibfield  {author} {\bibinfo {author} {\bibfnamefont {A.}~\bibnamefont
  {Dr{\'e}au}}, \bibinfo {author} {\bibfnamefont {A.}~\bibnamefont
  {Tchebotareva}}, \bibinfo {author} {\bibfnamefont {A.~E.}\ \bibnamefont
  {Mahdaoui}}, \bibinfo {author} {\bibfnamefont {C.}~\bibnamefont {Bonato}}, \
  and\ \bibinfo {author} {\bibfnamefont {R.}~\bibnamefont {Hanson}},\
  }\href@noop {} {\bibfield  {journal} {\bibinfo  {journal} {Physical Review
  Applied}\ }\textbf {\bibinfo {volume} {9}},\ \bibinfo {pages} {064031}
  (\bibinfo {year} {2018})}\BibitemShut {NoStop}%
\bibitem [{\citenamefont {Slattery}\ \emph {et~al.}(2019)\citenamefont
  {Slattery}, \citenamefont {Ma}, \citenamefont {Zong},\ and\ \citenamefont
  {Tang}}]{slattery2019background}%
  \BibitemOpen
  \bibfield  {author} {\bibinfo {author} {\bibfnamefont {O.}~\bibnamefont
  {Slattery}}, \bibinfo {author} {\bibfnamefont {L.}~\bibnamefont {Ma}},
  \bibinfo {author} {\bibfnamefont {K.}~\bibnamefont {Zong}}, \ and\ \bibinfo
  {author} {\bibfnamefont {X.}~\bibnamefont {Tang}},\ }\href@noop {} {\bibfield
   {journal} {\bibinfo  {journal} {Journal of Research of the National
  Institute of Standards and Technology}\ }\textbf {\bibinfo {volume} {124}},\
  \bibinfo {pages} {1} (\bibinfo {year} {2019})}\BibitemShut {NoStop}%
\bibitem [{\citenamefont {Ikuta}\ \emph {et~al.}(2019)\citenamefont {Ikuta},
  \citenamefont {Tani}, \citenamefont {Ishizaki}, \citenamefont {Miki},
  \citenamefont {Yabuno}, \citenamefont {Terai}, \citenamefont {Imoto},\ and\
  \citenamefont {Yamamoto}}]{ikuta2019frequency}%
  \BibitemOpen
  \bibfield  {author} {\bibinfo {author} {\bibfnamefont {R.}~\bibnamefont
  {Ikuta}}, \bibinfo {author} {\bibfnamefont {R.}~\bibnamefont {Tani}},
  \bibinfo {author} {\bibfnamefont {M.}~\bibnamefont {Ishizaki}}, \bibinfo
  {author} {\bibfnamefont {S.}~\bibnamefont {Miki}}, \bibinfo {author}
  {\bibfnamefont {M.}~\bibnamefont {Yabuno}}, \bibinfo {author} {\bibfnamefont
  {H.}~\bibnamefont {Terai}}, \bibinfo {author} {\bibfnamefont
  {N.}~\bibnamefont {Imoto}}, \ and\ \bibinfo {author} {\bibfnamefont
  {T.}~\bibnamefont {Yamamoto}},\ }\href@noop {} {\bibfield  {journal}
  {\bibinfo  {journal} {Physical review letters}\ }\textbf {\bibinfo {volume}
  {123}},\ \bibinfo {pages} {193603} (\bibinfo {year} {2019})}\BibitemShut
  {NoStop}%
\bibitem [{\citenamefont {Jeronimo-Moreno}\ \emph {et~al.}(2010)\citenamefont
  {Jeronimo-Moreno}, \citenamefont {Rodriguez-Benavides},\ and\ \citenamefont
  {U’Ren}}]{jeronimo2010theory}%
  \BibitemOpen
  \bibfield  {author} {\bibinfo {author} {\bibfnamefont {Y.}~\bibnamefont
  {Jeronimo-Moreno}}, \bibinfo {author} {\bibfnamefont {S.}~\bibnamefont
  {Rodriguez-Benavides}}, \ and\ \bibinfo {author} {\bibfnamefont {A.~B.}\
  \bibnamefont {U’Ren}},\ }\href@noop {} {\bibfield  {journal} {\bibinfo
  {journal} {Laser Physics}\ }\textbf {\bibinfo {volume} {20}},\ \bibinfo
  {pages} {1221} (\bibinfo {year} {2010})}\BibitemShut {NoStop}%
\bibitem [{\citenamefont {Pomarico}\ \emph {et~al.}(2009)\citenamefont
  {Pomarico}, \citenamefont {Sanguinetti}, \citenamefont {Gisin}, \citenamefont
  {Thew}, \citenamefont {Zbinden}, \citenamefont {Schreiber}, \citenamefont
  {Thomas},\ and\ \citenamefont {Sohler}}]{pomarico2009waveguide}%
  \BibitemOpen
  \bibfield  {author} {\bibinfo {author} {\bibfnamefont {E.}~\bibnamefont
  {Pomarico}}, \bibinfo {author} {\bibfnamefont {B.}~\bibnamefont
  {Sanguinetti}}, \bibinfo {author} {\bibfnamefont {N.}~\bibnamefont {Gisin}},
  \bibinfo {author} {\bibfnamefont {R.}~\bibnamefont {Thew}}, \bibinfo {author}
  {\bibfnamefont {H.}~\bibnamefont {Zbinden}}, \bibinfo {author} {\bibfnamefont
  {G.}~\bibnamefont {Schreiber}}, \bibinfo {author} {\bibfnamefont
  {A.}~\bibnamefont {Thomas}}, \ and\ \bibinfo {author} {\bibfnamefont
  {W.}~\bibnamefont {Sohler}},\ }\href@noop {} {\bibfield  {journal} {\bibinfo
  {journal} {New Journal of Physics}\ }\textbf {\bibinfo {volume} {11}},\
  \bibinfo {pages} {113042} (\bibinfo {year} {2009})}\BibitemShut {NoStop}%
\bibitem [{\citenamefont {Ikuta}\ \emph {et~al.}(2013)\citenamefont {Ikuta},
  \citenamefont {Kato}, \citenamefont {Kusaka}, \citenamefont {Miki},
  \citenamefont {Yamashita}, \citenamefont {Terai}, \citenamefont {Fujiwara},
  \citenamefont {Yamamoto}, \citenamefont {Koashi}, \citenamefont {Sasaki},
  \citenamefont {Wang},\ and\ \citenamefont {Imoto}}]{ikuta2013high-fidelity}%
  \BibitemOpen
  \bibfield  {author} {\bibinfo {author} {\bibfnamefont {R.}~\bibnamefont
  {Ikuta}}, \bibinfo {author} {\bibfnamefont {H.}~\bibnamefont {Kato}},
  \bibinfo {author} {\bibfnamefont {Y.}~\bibnamefont {Kusaka}}, \bibinfo
  {author} {\bibfnamefont {S.}~\bibnamefont {Miki}}, \bibinfo {author}
  {\bibfnamefont {T.}~\bibnamefont {Yamashita}}, \bibinfo {author}
  {\bibfnamefont {H.}~\bibnamefont {Terai}}, \bibinfo {author} {\bibfnamefont
  {M.}~\bibnamefont {Fujiwara}}, \bibinfo {author} {\bibfnamefont
  {T.}~\bibnamefont {Yamamoto}}, \bibinfo {author} {\bibfnamefont
  {M.}~\bibnamefont {Koashi}}, \bibinfo {author} {\bibfnamefont
  {M.}~\bibnamefont {Sasaki}}, \bibinfo {author} {\bibfnamefont
  {Z.}~\bibnamefont {Wang}}, \ and\ \bibinfo {author} {\bibfnamefont
  {N.}~\bibnamefont {Imoto}},\ }\href {\doibase 10.1103/PhysRevA.87.010301}
  {\bibfield  {journal} {\bibinfo  {journal} {Phys. Rev. A}\ }\textbf {\bibinfo
  {volume} {87}},\ \bibinfo {pages} {010301} (\bibinfo {year}
  {2013})}\BibitemShut {NoStop}%
\end{thebibliography}%

\end{document}